\documentclass[11pt,notitlepage]{article}
\usepackage[margin=1.2in,footskip=0.25in]{geometry}
\usepackage[parfill]{parskip}   
\usepackage[nodisplayskipstretch]{setspace}
\usepackage{enumerate}
\usepackage{bm}
\usepackage{amssymb}
\usepackage{amsmath}

\usepackage{amsthm}
\usepackage{graphicx}
\usepackage{caption}
\usepackage{subcaption}
\usepackage{makeidx}
\usepackage{multicol}
\usepackage{comment}
\usepackage{url}
\usepackage{color}
\usepackage[table]{xcolor}
\usepackage{chngcntr}
\usepackage{listings}
\usepackage{xcolor}
\definecolor{Gray}{gray}{0.9}
\newcolumntype{?}{!{\vrule width 2pt}}

\date{}

\begin{document}

\title{\vspace{-50pt}
\large\bf Quantum Buffer Design Using Petri Nets}

\author{\vspace{-35pt}\\ {\bf Syed Asad Shah } \\ Department of Computer Engineering \\ Bilkent University, Ankara, Turkey 06800 
\and {\bf  A. Yavuz Oru\c{c}} \\ Department of Electrical and Computer Engineering \\ University of Maryland, College Park, MD 20742}

\maketitle
\nopagebreak[4]
\setcounter{page}{1}

\vspace{-35pt}
\begin{center}
{\bf Abstract}
\end{center}

\vspace{-10pt}\noindent
This paper introduces a simplified quantum Petri net (QPN) model  and uses this model to generalize classical SISO, SIMO, MISO, MIMO and Priority buffers to their quantum counterparts. It provides a primitive storage element, namely a quantum S-R flip-flop and describes two different  such flip-flop designs using quantum NOT,  CNOT, CCNOT and SWAP gates. Each of the quantum S-R flip-flops can be replicated to obtain a quantum register for any given number of qubits. The aforementioned quantum buffers are then obtained using the simplified QPN model and quantum registers.  $\!\!$The quantum S-R flip-flop and quantum buffer designs have been tested using OpenQasm 2.0 and Qiskit programs on IBM quantum computers and simulators and the results validate their expected operations.
\vspace{-10pt}

\section{Introduction}
\label{problemStatement}

\vspace{-8pt}\noindent
Quantum computing and information processing research has evolved over the last four decades as a viable and persistent field of interest for computer scientists, engineers, physicists and even some mathematicians\cite{Lanzagorta2022,Djordjevic2021,Hassija2020,Horowitz-2019,Wolfgang2019}. What makes quantum computing and information processing attractive is the promise of quantum mechanics to provide nearly unlimited amount of parallelism in the smallest scales of physics that classical physical systems fail to offer without replicating computational resources. It is true that quantum parallelism is not suitable to handle every complex computational task but there is a sufficient set of computational problems, including factorization of large numbers, ordinary and constrained search and optimization problems, where quantum parallelism can be put to use to obtain solutions without using an exponentially increasing number of processing resources with the size of the problem in question \cite{Shor-1994,Grover-1996,Rhonda-2023}. There are several subfields of interest that are actively pursued in quantum computing and information processing research\cite{Preskill-2023,Humble-2021,Bacon-2010,Ramezani-2020,Aumasson-2017}. 

\vspace{-10pt}
\indent This paper focuses on the design of quantum buffers in which classical packets (bits) are replaced by quantum packets (qubits). Buffers play a fundamental role as temporary storage elements for efficient processing of information and to synchronize the information flow between various parts of classical computer and communications systems. Consequently, their design, implementation, and performance analysis have been extensively researched in the literature\cite{Danzig-1989,Kimura-1996,Kougkas-2020,Kim2001}.  As the paradigm of computing shifts from the classical to quantum computing, conventional buffering concepts should be transformed into quantum buffering concepts in order to pave the way for the design of future quantum computer and communication systems. Such an effort is timely as experimental quantum memory devices have been introduced recently and suggest that reliable quantum memories will likely be available  within the decade (See \cite{liu2023quantum} and the extensive list of research articles cited there).  
To explore the utility of such devices in quantum buffering, one first needs a solid model to describe the behavior of classical buffering and Petri nets  fulfill this need in both theoretical and practical terms \cite{Peterson-1977,Zurawski-1994,Zhou-1996}. A classical Petri net is an abstract model that describes the operations of asynchronous computational systems, where {\em tokens} in {\em places} initiate (fire) {\em transitions} to form other tokens when the transition requirements are met\cite{Peterson-1977}. Such requirements are usually characterized by the number of tokens entering a transition from neighboring places.  The operations of various types of buffers have been modeled using Petri nets to characterize their boundedness, liveness, safeness, and stability properties and to design reliable data processing systems\cite{Ye-2003}. Recently, classical Petri nets have been extended to quantum Petri nets in which  classical tokens are replaced by quantum tokens and classical transitions are replaced by quantum transitions using quantum gates\cite{Letia-2021,Letia-2022,Papavarnavas-2021}. The main objective of our work is to simplify this quantum Petri net model and use the simplified model to design quantum buffers. Quantum buffers have only been investigated in physical layers so far using fiber delay line methods\cite{Lee-2024}.  Our approach is concerned with a more theoretical and design aspect of quantum buffers, where we focus on the design of quantum flip-flops, registers and buffers, rather than implementing the storage of qubits using fiber spooling and other similar delay line techniques. Our quantum buffer designs can be viewed as quantum switching networks with buffering. In this setting, quantum packets are transferred from one buffer to another using quantum tokens that hold quantum information and quantum circuits that serve as transitions that are fired when they receive the required quantum tokens. Overall, we expect that quantum buffers will serve as temporary storage spaces  in quantum computers just like ordinary buffers in classical computers.  

\vspace{-12pt}\indent 
The rest of the paper is organized as follows. The next section presents out quantum Petri net  (QPN) model and provides two quantum S-R flip-flop designs that can be used to build quantum registers of a desirable size. Section 3 describes the designs of two quantum S-R flip-flops and Section 4 presents quantum SISO, SIMO, MISO, MIMO and priority  buffers. Section 5 provides examples and validations of our quantum buffer designs on IBM quantum computers and simulators. The paper is concluded in Section 6.

\par\noindent
\vspace{-35pt}
\section{The Quantum Petri Net Model}
\label{sectionpnk}

\vspace{-5pt}
The quantum Petri net $(QPN)$ model is a theoretical framework that extends the classical Petri nets to incorporate quantum mechanical phenomena into the model. The $QPN$ is designed to handle the behavior of quantum states, their evolutions, and probabilistic nature of quantum mechanics in a Petri net setting.  Two such models have been reported in the literature \cite{Papavarnavas-2021, Schmidt-2021}. The one that is used here is derived from the model introduced in \cite{Papavarnavas-2021} with some simplifications.

\begin{figure}
\vspace{-30pt}
\begin{center}
\begin{tabular}{c}
\includegraphics[height= 4.8 cm, width=\textwidth]{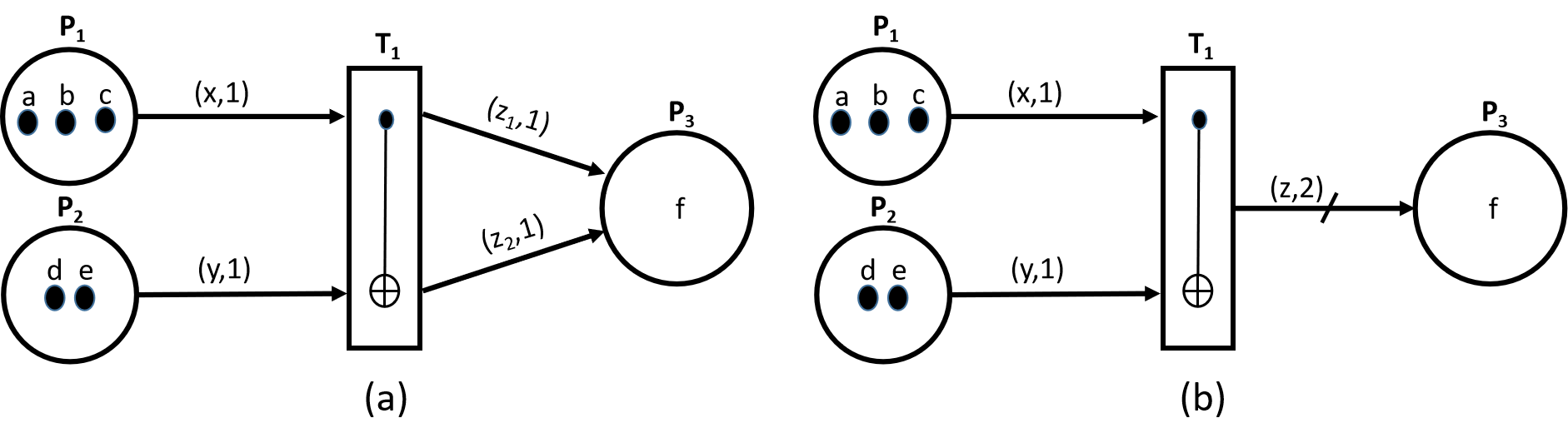}
\end{tabular}

\end{center}
\vspace{-5pt}
\caption 
{
(a) A QPN example with multiple arcs (b) A QPN example with compact representation of multiple arcs.} 
 \label{fig:spn}
\end{figure}

\vspace{-10pt}\noindent
\textbf{Definition}: A Quantum Petri net $(QPN)$ is a \text{6-tuple} $(D, P, T, E,  \mu, v)$ where: 

\vspace{-10pt}
\begin{enumerate}
\item $D$ is a finite set of quantum tokens $\{d_1, d_2, ..., d_r\}, r \geq 1,$  henceforth to be referred to as $q$-tokens, 
\vspace{-10pt}
\item $P$ is a finite set of places $\{P_1, P_2, P_3, \ldots, P_m\}$, $ m \geq 1, $
\vspace{-10pt}
\item $T$ is a finite set of transitions $\{T_1, T_2, T_3, \ldots, T_n\}$, $ n \geq 1, $ 
\vspace{-10pt}
\item $E$ is a finite set of directed and labeled arcs that connect places in $P$ with transitions in $T,$  where labels denote the $q$-tokens and their quantities for a transition to consume and fire or a place to acquire,
\vspace{-10pt}
\item $\mu,$   called a marking, is a mapping from $D$ to $P$, which assigns $q$-tokens to places,
\vspace{-10pt}
\item \textit{$v(P_i,\mu, t)$} is an assignment of qubits to $q$-tokens in place $P_i$  in marking $\mu$ at time $t = 0, 1, 2, ... .$
\end{enumerate}
The $q$-tokens in quantum Petri nets represent a set of quantum bits (qubits) or multi-qubits.  Qubits are the basic units of quantum information. The $q$-tokens are represented by small black-filled circles in QPN diagrams as shown in Figure~\ref{fig:spn}, where they are identified by letters $a, b, c, d, e,$ etc. Places are locations, where $q$-tokens reside.  They are represented by the large hollow circles in the figure.  Each place may hold a certain number of $q$-tokens, and the state of a place is determined by  the  $q$-tokens it contains. Places and $q$-tokens collectively represent the overall state of a QPN. Transitions represent  quantum  events (operations) that can change the state of a QPN, and manipulate $q$-tokens using quantum gates \cite{Brylinski-2002}. Directed arcs 
 in $E$ link places and transitions together.  The direction of an arc determines the flow of $q$-tokens. An arc from a place to a transition indicates that tokens in that place can be consumed by that transition when it fires. An arc from a transition to a place indicates that tokens can be added to that place when the transition fires. As seen in Figure~\ref{fig:spn}(a), multiple $q$-tokens can be added to a place at once. Figure~\ref{fig:spn}(b) shows a more compact representation of the QPN in Figure~\ref{fig:spn}(a), where the output arcs from transition $T_1$ are fused together to a single output arc with a label of $(z,2)$, indicating that the arc carries two tokens and $z = (z_1,z_2).$ Henceforth, we will use the compact QPN representation. The labels on the incoming arcs to a transition such as $x$ and $y$ denote variables for $q$-tokens that are fed into a transition in the indicated quantities. Similarly the labels on the outgoing arcs from a transition such as $z$ denote variables for $q$-tokens that are added into a place in the indicated quantities. For example, one $q$-token in each of places $P_1$ and  $P_2$ ($x = a$ and $y = d$) are used by transition $T_1$ in Figure~\ref{fig:spn}, which in this case is a controlled-not (C-NOT) gate. It is assumed that $q$-tokens are consumed by transitions in some predetermined order.  In the examples presented here, an alphabetical order will be used to name tokens unless otherwise stated. When fired, a transition manipulates one or more $q$-tokens according to the function that is specified, and moves the generated $q$-tokens to the output places that are connected to that transition. For example, in Figure~\ref{fig:spn}, $z = (a, a\oplus d)$ is placed in $P_3,$ i.e., $f$ becomes $z$ after $T_1$ fires. Transitions in a QPN may represent quantum gates, measurements, or other quantum operations.

\vspace{-10pt}\indent
Fired transitions are the only means by which $q$-tokens (qubits) move along the arcs.  In  quantum computing context, these transitions of $q$-tokens in a QPN represent the transformation of quantum states and how they evolve due to quantum parallelism. Effectively, extending the classical Petri net model to a quantum Petri net model adds another layer of control over quantum circuits, one in which quantum operations can be made conditional on the quantities and content of $q$-tokens. 

\vspace{-10pt}\indent
A marking in a QPN defines the distribution of tokens across the places that represent various configurations or states of the quantum system. The initial state of a QPN is specified by its marking that assigns $q$-tokens to its places. The evolution of the system is then represented by a sequence of changes in the states of the places, driven by the firing of transitions. As transitions fire and tokens move, they alter the overall quantum state of the system. The function $v(P_i,\mu, t)$ represents an assignment of qubits to  $q$-tokens under the marking $\mu$ at time $t$ in place $P_i$. Subsequent changes evolve with every transition executed, resulting in $v(P_i,\mu, 0), v(P_i,\mu, 1),\ldots$ The following example illustrate these ideas.

\noindent
{\bf Example:}

\vspace{-10pt}\noindent
As can be seen in Figure~\ref{fig:spn}, places $P_1$, $P_2$, and $P_3$ have $q$-tokens assigned to them under the following map $\mu,$ where $\mu(a) = \mu(b) = \mu (c)  = P_1,$ and  $\mu(d) = \mu(e)  = P_2$. For this QPN to operate we initialize $q$-tokens to qubits as follows: 

\vspace{-25pt}
\begin{align*}
v(P_1,\mu, 0) &= 
\left\{
\begin{array}{l}
a: |1\rangle \\ 
b: |1\rangle \\ 
c: \frac{1}{\sqrt{2}}(|0\rangle + |1\rangle)
\end{array} 
\right\}, \\
v(P_2,\mu, 0) &= 
\left\{
\begin{array}{l}
d: |0\rangle \\
e: |1\rangle
\end{array}
\right\}.
\end{align*}
With these assignments, transition $T_1$ will fire at $t = 0,$ and  the places $P_1$, $P_2$, and $P_3$ will have the following $q$-tokens at $t = 1,$  where $T_1$ inverts $d$ qubit to $|1\!\!>:$

\vspace{-20pt}
\begin{align*}
v(P_1,\mu, 1) &= 
\left\{
\begin{array}{l}
b: |1\rangle \\ 
c: \frac{1}{\sqrt{2}}(|0\rangle + |1\rangle)
\end{array} 
\right\}, \\
v(P_2,\mu, 1) &= 
\left\{
\begin{array}{l}
e: |1\rangle
\end{array}
\right\}, \\
v(P_3,\mu, 1) &= 
\left\{
\begin{array}{l}
a: |1\rangle
\\
d: |1\rangle
\end{array}
\right\}. \qed
\end{align*}

\begin{figure}
\vspace{-20pt}
\begin{center}
\begin{tabular}{c}
\includegraphics[height= 4.8 cm]{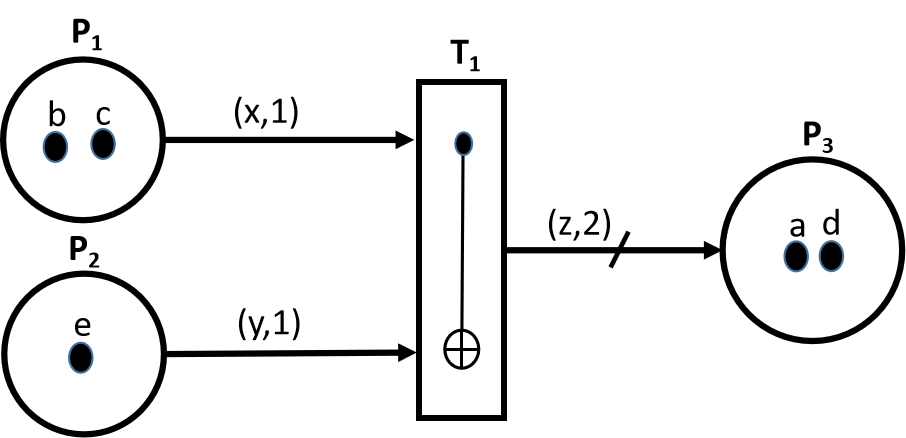}
\end{tabular}
\end{center}
\vspace{-5pt}
\caption 
{ 
A transition $T_1$ fires, at time $t=0.$ } 
\label{fig:spn1}
\end{figure}

\vspace{-5pt}\noindent
The new state of the QPN is depicted in Figure~\ref{fig:spn1} with the $q$-tokens $a$ and $d$ moved into place $P_3$  with $d$ flipped after the transition. In the rest of the paper, we will employ the QPN model to design four different quantum buffers.

$\,$\\

\vspace{5pt}

\vspace{-40pt}
\section{Quantum  Flip-Flops}
\label{quantumFlip-Flops}

\vspace{-8pt}
For the purposes of this paper, a quantum buffer is a quantum system in which collections of qubits are structured into quantum packets, which are  moved from a set of inputs to a set of outputs.  Unlike classical buffers that store  and process collections of  bits  (0's and 1's), quantum buffers must obey the properties of superposition, entanglement and coherence of quantum systems. The physical designs of quantum memories using various physical models have been described in~\cite{Simon-2010}. Here, we focus on the conceptual design of quantum buffers using a primitive building block, called a quantum S-R (Q-S-R) flip-flop. Quantum flip-flops are similar to classical flip-flops except that they store quantum bits rather than a classical bit.  We note that quantum D, T, and J-K flip-flops have been reported earlier in the literature~\cite{misra-2018,misra-2024} using three quantum circuits, called the R-CQCA, PPM,  and F2G  gates. Both R-CQCA and PPM circuits have four inputs and four outputs and can be used to create a number of quantum gates as well as design quantum flip-flops. The F2G gate has three inputs and three outputs and is used in two different designs of D, T, and J-K flip-flops.  
We refer the reader to \cite{misra-2018,misra-2024} for  a detailed description of these quantum flip-flop designs. We will compare the Q-S-R flip-flops described in this paper with the two quantum J-K flip-flop designs described in ~\cite{misra-2018,misra-2024} in Section~\ref{runs} as a Q-S-R flip-flop can be obtained from a Q-J-K flip-flop rather easily by limiting the J and K inputs to $|00\rangle, |01\rangle,$ and $|10\rangle$ inputs.  

\begin{figure}
\vspace{-20pt}
\begin{center}
\begin{tabular}{c}
\includegraphics[height= 4.5 cm, width=11 cm]{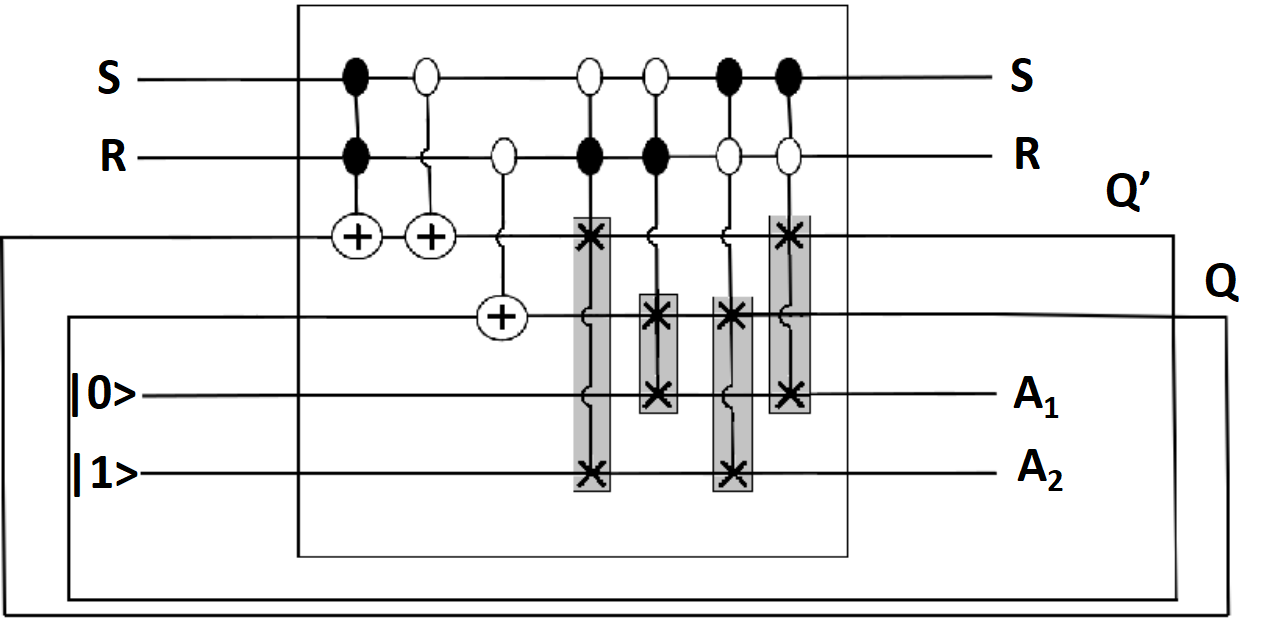}
\end{tabular}
\end{center}
\vspace{-5pt}
\caption 
{ 
 A Quantum S-R (Q-S-R) Flip-Flop Circuit 1. } 
\label{fig:qsr1}
\end{figure} 
\vspace{10pt}
\begin{figure}
\vspace{0 pt}
\begin{center}
\begin{tabular}{c}
\includegraphics[height= 4.3 cm]{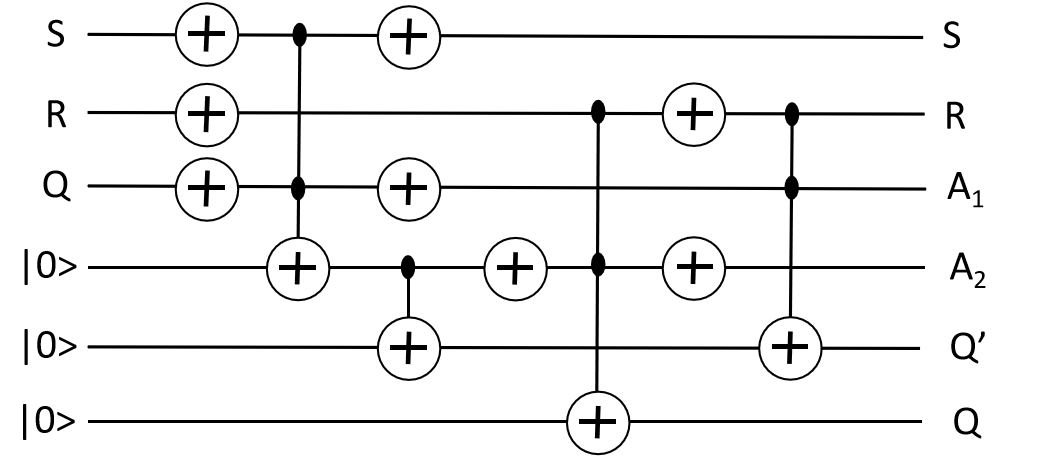}
\end{tabular}
\end{center}
\vspace{-5pt}
\caption 
{ 
 A Quantum S-R (Q-S-R) Flip-Flop Circuit 2. } 
\label{fig:qsr2}
\end{figure} 

 \vspace{-20pt}
Our designs of a Q-S-R flip-flop are depicted in Figure~\ref{fig:qsr1} and Figure~\ref{fig:qsr2} , where \( |0\rangle \) and \( |1\rangle \) are used as ancillary qubits to set  and reset the $Q$ output. The Q-S-R flip-flop in Figure~\ref{fig:qsr1} has been designed using two C-NOT, one CC-NOT and four controlled quantum swap gates with two control inputs $S$ and $R.$ The first two controlled swap gates swap their inputs when $S = |0\!\!>$ and $R = |1\!\!>,$ whereas the last two controlled swap gates swap their inputs when $S = |1\!\!>$ and $R = |0\!\!>\!\!.$ The second Q-S-R flip-flop has been designed using  one C-NOT, three CC-NOT gates and  eight quantum NOT gates. The next state behavior of a Q-S-R flip-flop is described in Table~\ref{tab:srqTable}. By tracing the qubits from the inputs to the outputs, it can be verified  that  Circuit 1 and Circuit 2 both  behave like a Q-S-R flip-flop as defined in Table~\ref{tab:srqTable} when their  $S$ and $R$ inputs are set to $|00\rangle, |01\rangle, |10\rangle.$ 
On the other hand,  if $S$ and $R$ are set to $|11\rangle$ in the Q-S-R flip-flop in Figure~\ref{fig:qsr1}, then the outputs $Q$ and $Q'$ will both be forced to $|1\rangle$ if $Q = |0\rangle$ and to $|0\rangle$ if $Q = |1\rangle$, violating the assumption that $Q$ and $Q'$ are complements of each other. More importantly,  if the $S$ and $R$ inputs return to 0 about the same time  then the outputs $Q$ and $Q'$  may race to qubit $|1\rangle$ that may then potentially lead to an undefined state as in a classical S-R flip-flop. Similarly, in the Q-S-R flip-flop in Figure~\ref{fig:qsr2}, if $S$ and $R$ are $|11\rangle$ then both outputs $Q$ and $Q'$  will be forced to $|0\rangle$, contradicting the assumption that $Q$ and $Q'$ oppose each other. As in the earlier case, when $S$ and $R$ inputs both go to $|0\rangle$ about the same time, the outputs $Q$ and $Q'$ may race to qubit $|1\rangle.$

\vspace{10pt}
\begin{table}[ht] 
 {\small   \centering
    \begin{tabular}{|c|c|c|c?c|} 
    \hline
    S & R & Q &Q$' $&Q-Output \\ 
    \hline
    \( |0\rangle \) & \( |0\rangle \) & \( |0\rangle \) &\( |1\rangle \) & \( |0\rangle \) \\ 
    \hline
   \( |0\rangle \) & \( |0\rangle \) & \( |1\rangle \) &\( |0\rangle \) & \( |1\rangle \) \\ 
    \hline
    \( |1\rangle \) & \( |0\rangle \) & \( |0\rangle \) &\( |1\rangle \) &  \( |1\rangle \) \\ 
    \hline
    \( |1\rangle \) & \( |0\rangle \) & \( |1\rangle \) &\( |0\rangle \) &  \( |1\rangle \) \\ 
    \hline
    \( |0\rangle \) & \( |1\rangle \) & \( |0\rangle \) &\( |1\rangle \) & \( |0\rangle \) \\ 
    \hline
    \( |0\rangle \) & \( |1\rangle \) & \( |1\rangle \) &\( |0\rangle \) & \( |0\rangle \) \\ 
    \hline
    \( |1\rangle \) & \( |1\rangle \) & \( |0\rangle \)&  \( |1\rangle \) & Undefined \\ 
     \hline
    \( |1\rangle \) & \( |1\rangle \) & \( |1\rangle \) & \( |0\rangle \) & Undefined \\ 
    \hline  
 \end{tabular}

 \vspace{1pt}
    \caption{The next state behavior of  a Q-S-R flip-flop.}
    \label{tab:srqTable} 
}\end{table}

\begin{figure}
\vspace{-20pt}
\begin{center}
\begin{tabular}{c}
\includegraphics[width=13cm]{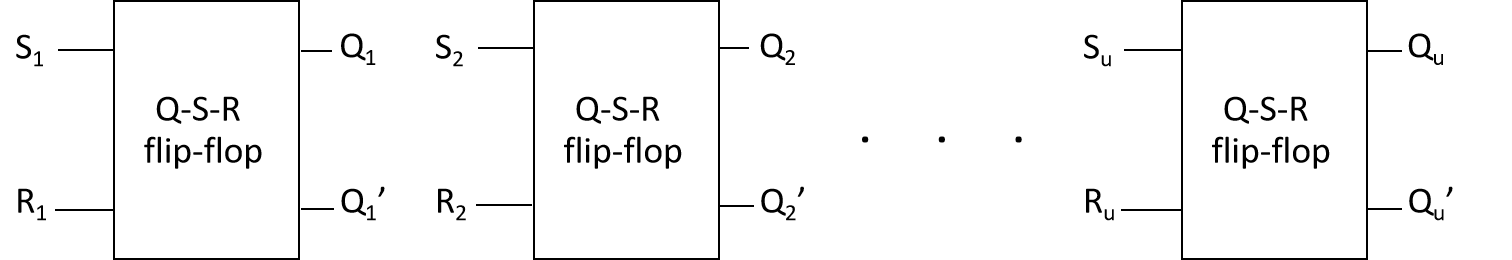}
\end{tabular}
\end{center}
\vspace{-5pt}
\caption 
{ 
A $u$-qubit quantum register, constructed out of $u$ Q-S-R flip-flops.} 
\label{fig:qsrn}
\end{figure}

\vspace{-5pt}
\section{Quantum Buffers}

\vspace{-4pt}
\label{quantumbuffering}

\vspace{-5pt}\noindent
Quantum registers of a desirable size can be obtained using Q-S-R flip-flops as in the design of classical registers. For example, a $u$-qubit  register can be put together using $u$ Q-S-R flip-flops as shown in Figure~\ref{fig:qsrn}. 
Quantum registers are building blocks of quantum buffers that store and process collections of qubits using quantum gates and circuits.  The QPN model that is introduced in Section~\ref{sectionpnk} facilitates this quantum processing, where quantum registers correspond to the places in the QPN model.  Places can hold data $q$-tokens (multi-qubits) or ancillary $q$-tokens (multi-qubits). The ancillary $q$-tokens may represent single qubits  in some buffer designs and multi-qubits in other buffer designs. They are used used to limit the capacity of a quantum buffer or select a particular transition to fire. 
The places that exclusively hold data $q$-tokens will be denoted by $P_D,$ those that exclusively hold ancillary $q$-tokens will be denoted by $P_A,$ and those that hold both data ancillary $q$-tokens  will be denoted by $P_{DA}$ with appropriate indexing, such as $P_{DA,1}, P_{DA,2},$ etc. The places from which $q$-tokens are fed into and taken out of the buffer will be denoted by $P_I$ and $P_O,$ respectively with appropriate indexing as well. 

\vspace{-10pt}
\subsection{Single Input/Single Output  Quantum Buffer}

\vspace{-4pt}\noindent
An SISO (Single Input, Single Output)  quantum buffer is a type of quantum buffer that has only one entry and one exit point for data $q$-tokens. The  SISO buffer is designed to transfer up to $m$ $q$-tokens from an input place $P_I$ of $n$ data $q$-tokens to an output place $P_O$  that can hold up to $m$ $q$-tokens, where it is assumed that $m\le n.$  The transfer of data $q$-tokens stops when the last ancillary qubit in $P_A$ is consumed by $T_1.$ The QPN model of the SISO is thus designed to have a capacity $m,$  and it consists of a single transition, $T_1$, four places  $P_I, P_A, P_{A1}, P_O$, $n$ data $q$-tokens  $d_1,d_2, ...,d_n,$ and $m$ ancillary $q$-bits $z_1,..., z_m,$ all of which are initialized to $|0\!\!>$ qubit. The transition $T_1$ is the identity quantum gate, denoted by $I$, that moves the $q$-tokens in $P_I$ to $P_O$ and the qubits in $P_A$ to $P_{A1} .$ The $n$ data  $q$-tokens are placed in $P_I$ and $m$ ancillary qubits are placed in $P_A$  before the buffer starts operating as follows:

\begin{figure}
\vspace{-30pt}
\begin{center}
\begin{tabular}{c}
\includegraphics[height= 4.6cm,width=8cm]{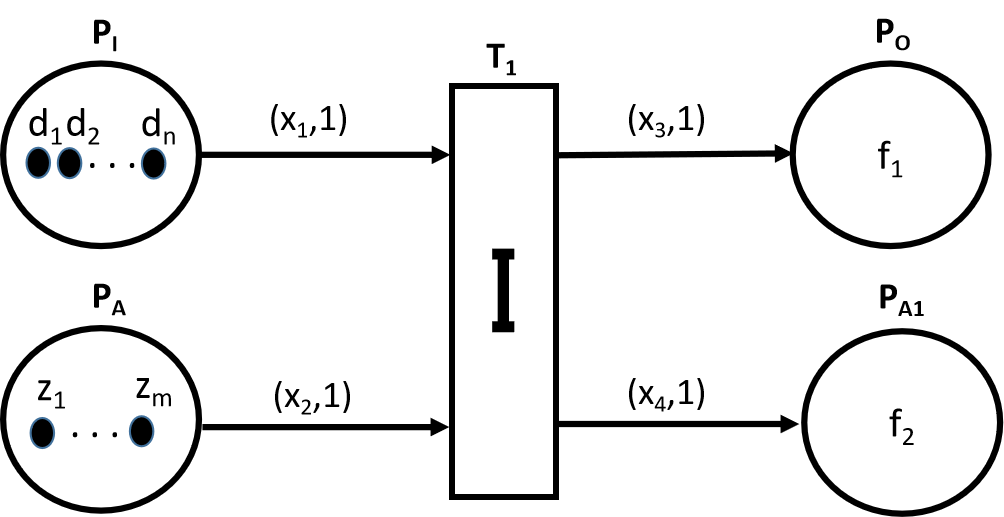}
\end{tabular}
\end{center}
\vspace{-5pt}
\caption 
{
A single-input-single-output (SISO) quantum buffer. (Note: The buffer has a single port to receive data $q$-tokens from place $P_1$ and a single port to output them to place $P_o.$ The ancillary $q$-tokens are there to limit the number of transfers $P_I$ to $P_o$ to $m$ $q$-tokens as the I-gate stops firing when all $m$ ancillary qubits are consumed.) }
 \label{fig:siso}
\end{figure}

\vspace{-35pt}
\begin{align*}
 \mu(d_i) &= P_I, 1\le i\le n,\\
 \mu(z_j) &= P_A, 1\le j\le m
\end{align*}

\vspace{-35pt}
\begin{align*}
v(P_I,\mu, 0) = 
\left\{
\begin{array}{c}
d_1: d^*_1 \\
d_2: d^*_2 \\
\cdot \\[-5pt]
\cdot \\[-5pt]
\cdot \\
d_n: d^*_n 
\end{array} 
\right\}, \quad
v(P_A,\mu, 0) = 
\left\{
\begin{array}{c}
z_1: |0\!> \\
z_2: |0\!> \\
\cdot \\[-5pt]
\cdot \\[-5pt]
\cdot \\
z_m: |0\!>
\end{array} 
\right\}.
\end{align*}

\begin{figure}
\vspace{-45pt}
\begin{center}
\begin{tabular}{c}
\includegraphics[height= 6.3 cm]{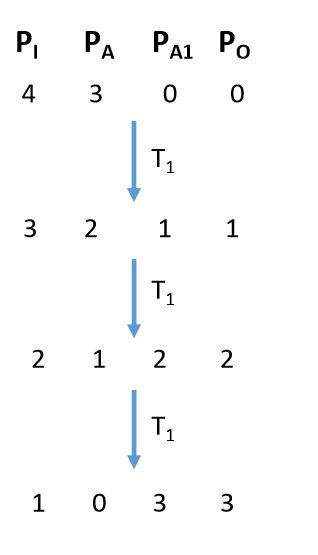}
\end{tabular}
\end{center}
\vspace{-15pt}
\caption 
{ 
The flow of $q$-tokens in a quantum SISO buffer, $n = 4, m = 3.$} 
\label{fig:sisoM}
\end{figure}

\vspace{-10pt}\noindent
Here, the variables with superscript $*$ denote the qubits or multi qubits that are assigned to the data $q$-tokens, $d_1,d_2,\cdots, d_n,$ at time $t = 0.$ When the input transition $T_1$ fires, it forwards the current data $q$-token and ancillary qubit, say $d_k$ and $z_k$ to places $P_O$ and $P_{A1}$ respectively i.e., $f_1$ becomes $d_k$ and $f_2$ becomes $z_k$. The behavior of the SISO buffer of capacity $m = 3$ is illustrated in Figure~\ref{fig:sisoM}.  The values that are  listed underneath the place labels denote the number of $q$-tokens  that are currently located in corresponding places. As seen at the top of  Figure~\ref{fig:sisoM}, $P_I, P_A, P_{A1}, P_O$ initially hold  4, 3, 0, and 0 $q$-tokens, and  the number of $q$-tokens in each place changes as the transition $T_1$ continues to fire.  The quantities of $q$-tokens at the end of each  transition can be verified by tracing the $q$-tokens as they move between places.  To process $m$ data $q$-tokens, $m$ steps (transitions) are required and the whole process is reversible, and can be repeated again.    

$\,$
\vspace{-30pt}
\subsection{Multiple Input/Single Output Quantum Buffer}
\begin{figure}
\vspace{-45pt}
\begin{center}
\begin{tabular}{c}
\includegraphics[height= 6 cm,width=9.5 cm]{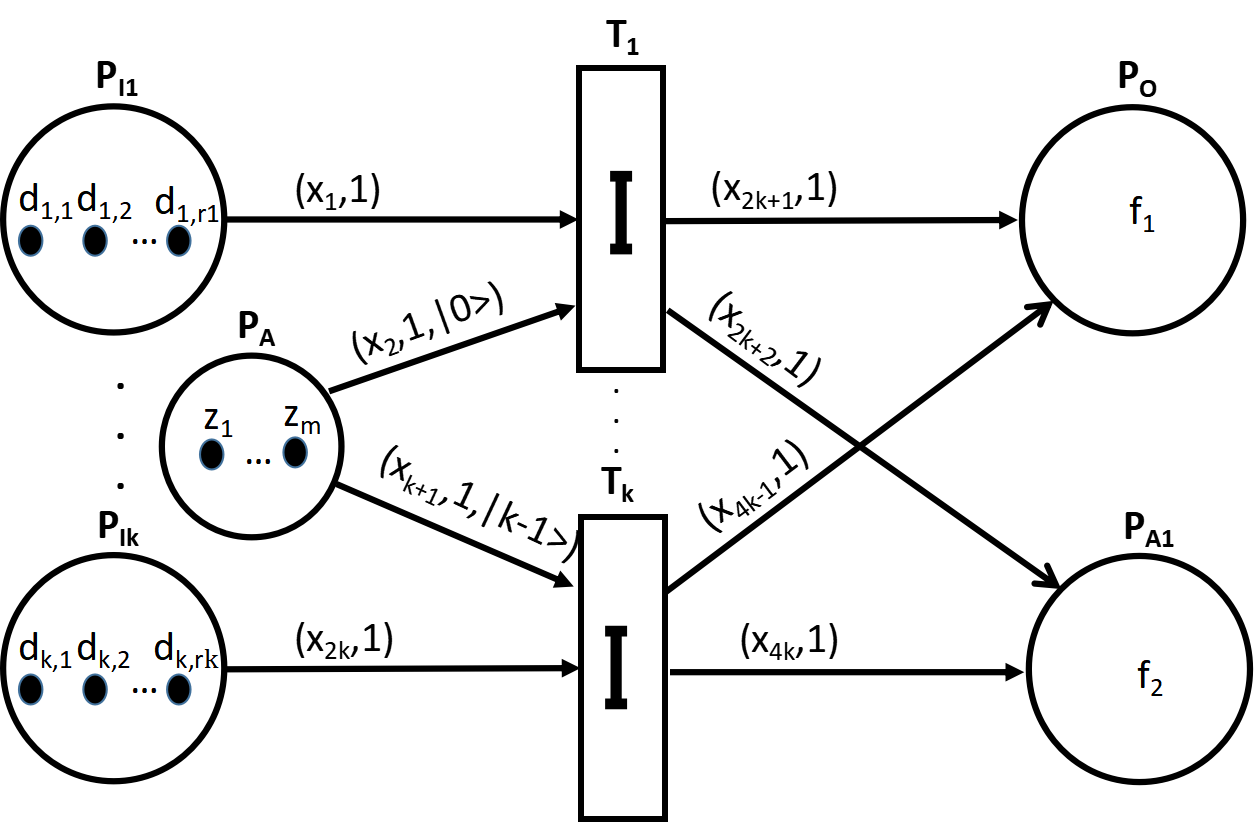}
\end{tabular}
\end{center}
\vspace{-5 pt}
\caption 
{ 
A multiple-input-single-output (MISO) quantum buffer. }
\label{fig:miso}
\end{figure}

\vspace{-2pt}\noindent
A MISO (Multiple Input, Single Output) quantum buffer is a type of quantum buffer that has many entries and only one exit point for data $q$-tokens. The MISO quantum buffer allows pairing any of its $k$ data inputs (places) with its only output  place to transfer a $q$-token from that input to the output. Figure~\ref{fig:miso} depicts the QPN model of MISO quantum buffer with capacity $m,$ $k$ transitions $T_1, T_2, ... T_{k}$, $k+3$ places $P_{I1}, P_{I2}, ... P_{Ik},  P_A, P_{A1}, P_O$, $r_j$ data $q$-tokens $d_{1,1}, d_{1,2}, ..., d_{j,r_j}$ in $P_{Ij}, 1\le j\le k,$ and $m$ ancillary $q$-tokens $z_1,..., z_m$. Each $z_i$ is specified as an address $q$-token that is used to select one of the transitions $T_1, T_2, ... T_{k}$ to fire in order to transfer a data $q$-token in one of the input places $P_{I1}, P_{I2}, ... P_{Ik}$ to output place $P_O.$ This is accomplished\footnote{Conditional transitions are permitted in quantum petri nets\cite{Papavarnavas-2021}.} by a condition placed on each link that compares the current $z_i,1\le i\le m$ token against the qubits that are marked on the arcs between $P_A$ and $T_i,1\le i \le k.$ Each of $z_1,..., z_m$ causes one transition to fire and the  MISO buffer halts its operation when all  $m$ address $q$-tokens are consumed.  The labels on the directed arcs $x_1, x_2,\cdots,x_{4k+4}$ denote the variables that represent data $q$-tokens or pairs of data and address $q$-tokens on the incoming and outgoing arcs of transitions. The  $q$-tokens are assigned to places $P_{I1},P_{I2},... P_{Ik},$ and $P_A$ are initialized before the buffer starts operating as follows:

\vspace{-30pt}
\begin{align*}
 \mu(d_{1,i}) = P_{I1},  1\le i\le r_1,
 \mu(d_{2,i}) = P_{I2},  1\le i\le r_2,
 &\cdots
 \mu(d_{k,i}) = P_{Ik},  1\le i\le r_k,\\
 \mu(z_q) = P_A,  1\le q\le m,
\end{align*}

\vspace{-30pt} 
\begin{align*}
v(P_{I1},\mu, 0) &= 
\left\{
\begin{array}{c}
d_{1,1} : d^*_{1,1} \\
d_{1,2} : d^*_{1,2}\\
\cdot \\ [-5pt]
\cdot \\ [-5pt]
\cdot \\
d_{1,r} : d^*_{1,r_1}
\end{array}
\right\}, 
\cdots,
v(P_{Ik},\mu, 0) = 
\left\{
\begin{array}{c}
d_{k,1} : d^*_{k,1}\\
d_{k,2} : d^*_{k,2}\\
\cdot \\ [-5pt]
\cdot \\ [-5pt]
\cdot \\
d_{k,r_k} : d^*_{k,r_k}
\end{array} 
\right\},  \\  \\
v(P_A,\mu, 0)& =
\left\{
\begin{array}{c}
z_1 : z*_1 \in \{|0>,|1>,\cdots, |k-1>\}\\
z_2 : z*_2 \in \{|0>,|1>,\cdots, |k-1>\}\\
\cdot \\ [-5pt]
\cdot \\ [-5pt]
\cdot \\
z_m : z*_{m} \in \{|0>,|1>,\cdots, |k-1>\}>
\end{array} 
\right\}.
\end{align*}

\begin{figure}
\vspace{-10pt}
\begin{center}
\begin{tabular}{c}
\includegraphics[height= 4cm, width=6cm]{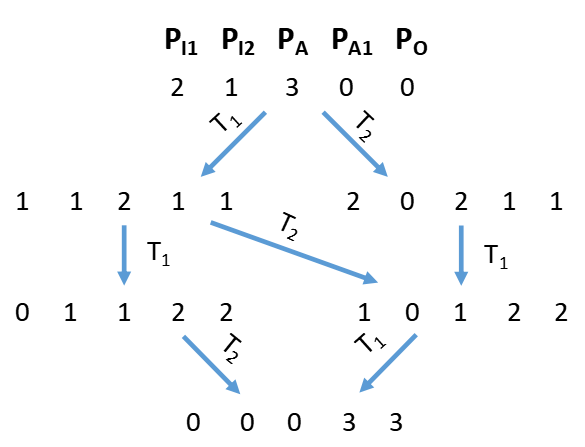}
\end{tabular}
\end{center}
\vspace{-10pt}
\caption 
{ 
The flow of $q$-tokens in a quantum MISO buffer, $k = 2, r_1 = 2, r_2 = 1, m = 3.$} 
\label{fig:misoM}
\end{figure}

\vspace{-10pt}
\noindent 
The behavior of the MISO quantum buffer of capacity $m = 2$ with two input places $P_{I1}$ and $P_{I2}$ is illustrated in Figure~\ref{fig:misoM}. The transitions are fired according to the address values of the ancillary qubits and the number of $q$-tokens in the input places. For example, transition $T_1$ can fire twice along a given sequence of transitions since $m = 2$ and place $P_{I1}$ contains two $q$-tokens. On the other hand, transition $T_2$ can only fire once as place $P_{I2}$ has only one $q$-token. The quantities of $q$-tokens at the end of each  transition can be verified by tracing the $q$-tokens as they move between places. Some transitions cannot be fired due to $q$-token constraints. To process $m$ data $q$-tokens, $m$ steps (transitions) are required, the whole process is reversible, and can be repeated again.

\vspace{-8pt}
\subsection{Single Input/Multiple Output Quantum Buffer}
\begin{figure}
\vspace{-30pt}
\begin{center}
\begin{tabular}{c}
\includegraphics[height= 6.4 cm,width= 8cm]{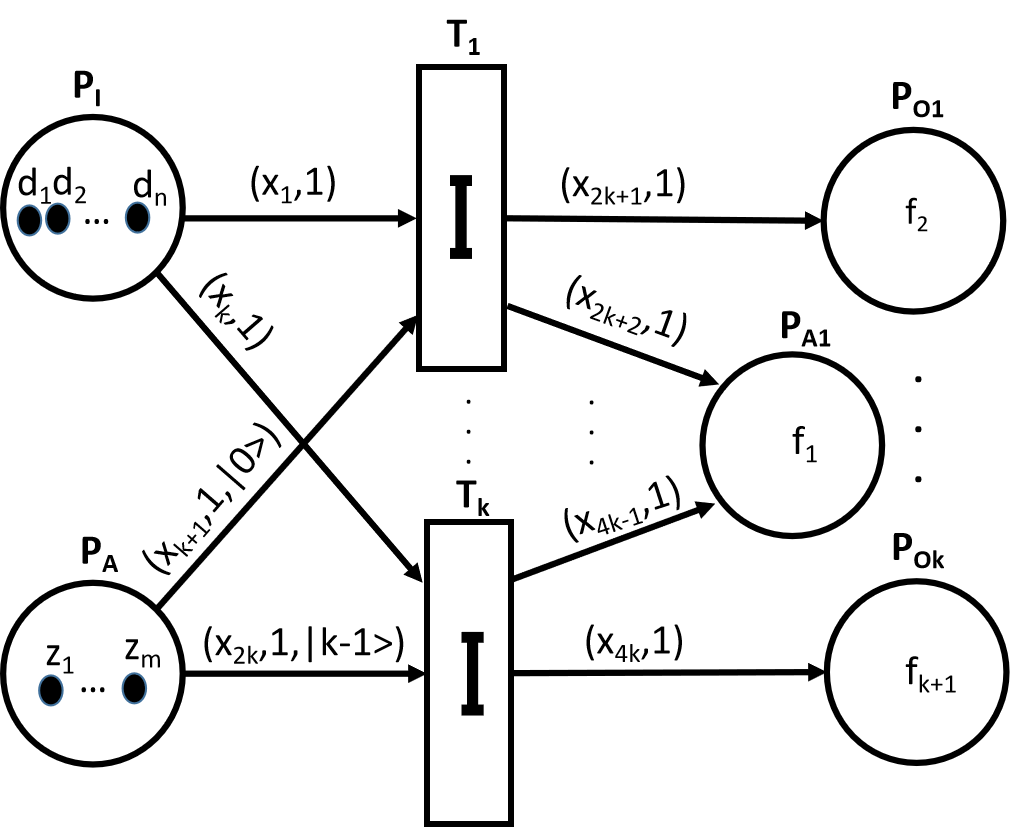}
\end{tabular}
\end{center}
\vspace{-5pt}
\caption 
{ 
A single-input-multiple-output (SIMO) quantum buffer.}
\label{fig:simo}
\end{figure}

\begin{figure}
\vspace{-10pt}
\begin{center}
\begin{tabular}{c}
\includegraphics[height= 5.3 cm,width=14cm]{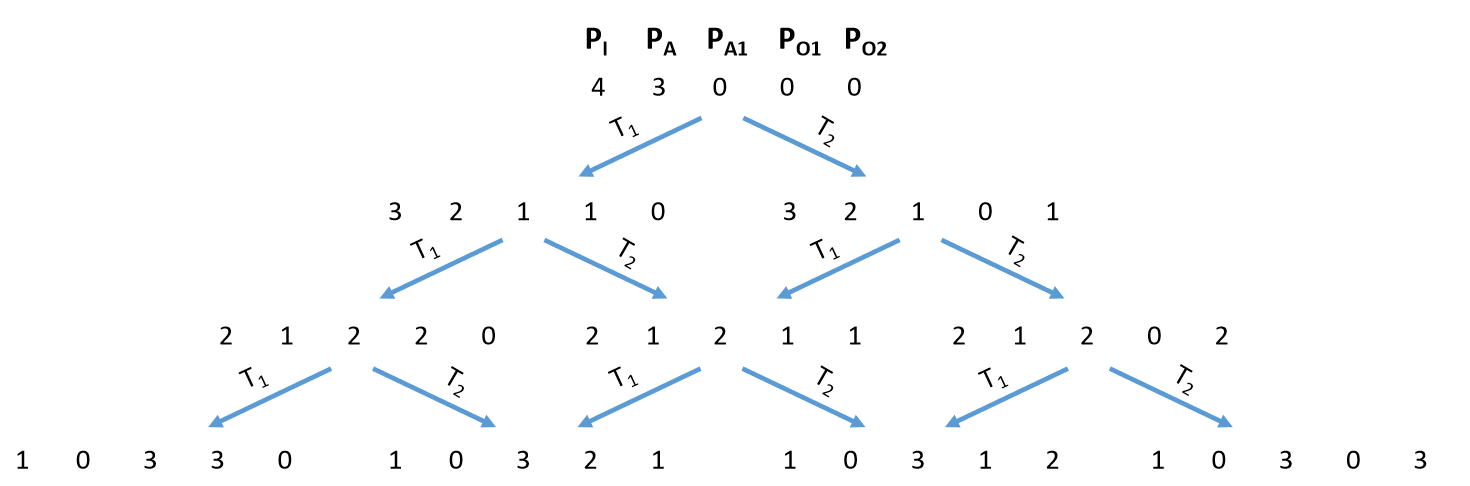}
\end{tabular}
\end{center}
\vspace{-10pt}
\caption 
{
The flow of $q$-tokens in a quantum SIMO buffer, $k = 2, m = 3, n\ge 3.$} 
 \label{fig:simoM}
\end{figure}

\vspace{-5pt}\noindent
A SIMO (Single Input, Multiple Output) quantum buffer is a type of quantum buffer that has only one entry and many exit points for data $q$-tokens.  Figure~\ref{fig:simo} depicts the QPN model of SIMO quantum buffer of capacity $m,$ composed of $k$ transitions $T_1,T_2, ... T_k$, $k+3$ places $P_I, P_A, P_{A1}, P_{O1}, P_{O2}, ... {P}_{Ok}$, $n$ data $q$-tokens $d_1,d_2, ..., d_n,$ and $m$ ancillary $q$-tokens $z_1,..., z_m,$ each of which can be assigned a quantum value from $|0\!>$ to $|k-1\!\!>$. The labels on the directed arcs $x_1, x_2,\cdots,x_{4k}$ denote the variables that represent $q$-tokens on the incoming and outgoing arcs of transitions.
The initial $q$-tokens are assigned to places $P_I$ and $P_A$ as in SISO quantum buffer. When transition $T_j, 1\le j\le k$ fires,  it forwards the current data and address $q$-tokens to $P_{Oj}$ and $P_{A1},$ respectively. The behavior of the SIMO buffer of capacity $m = 3$ with two output places $P_{O1}$ and $P_{O2}$ is illustrated in Figure~\ref{fig:simoM}. 
The quantities of $q$-tokens at the end of each  transition can be verified by tracing the $q$-tokens as they move between places.  We note that at the end of possible transitions, the four data $q$-tokens may be distributed to $P_{O1} $ and $P_{O2}$  in one of four ways: (3,0), (2,1), (1,2), (0,3) as the bottom row in Figure~\ref{fig:simoM} shows. Also, the value of $m$ limits the number of transitions to 3 in this example, preventing transitions with output patterns (4,0)  (3,1), (2,2), (1,3), (0,4). The diagram shows only the feasible sequences of transitions, and to process $m$ data $q$-tokens, $m$ steps (transitions) are needed, where the whole process is reversible, and can be repeated again. 
\vspace{-10pt}
\subsection{Multiple Input Multiple Output Quantum Buffer}
\begin{figure}
\vspace{-35pt}
\begin{center}
\begin{tabular}{c}
\includegraphics[height= 6.5cm,width=14cm]{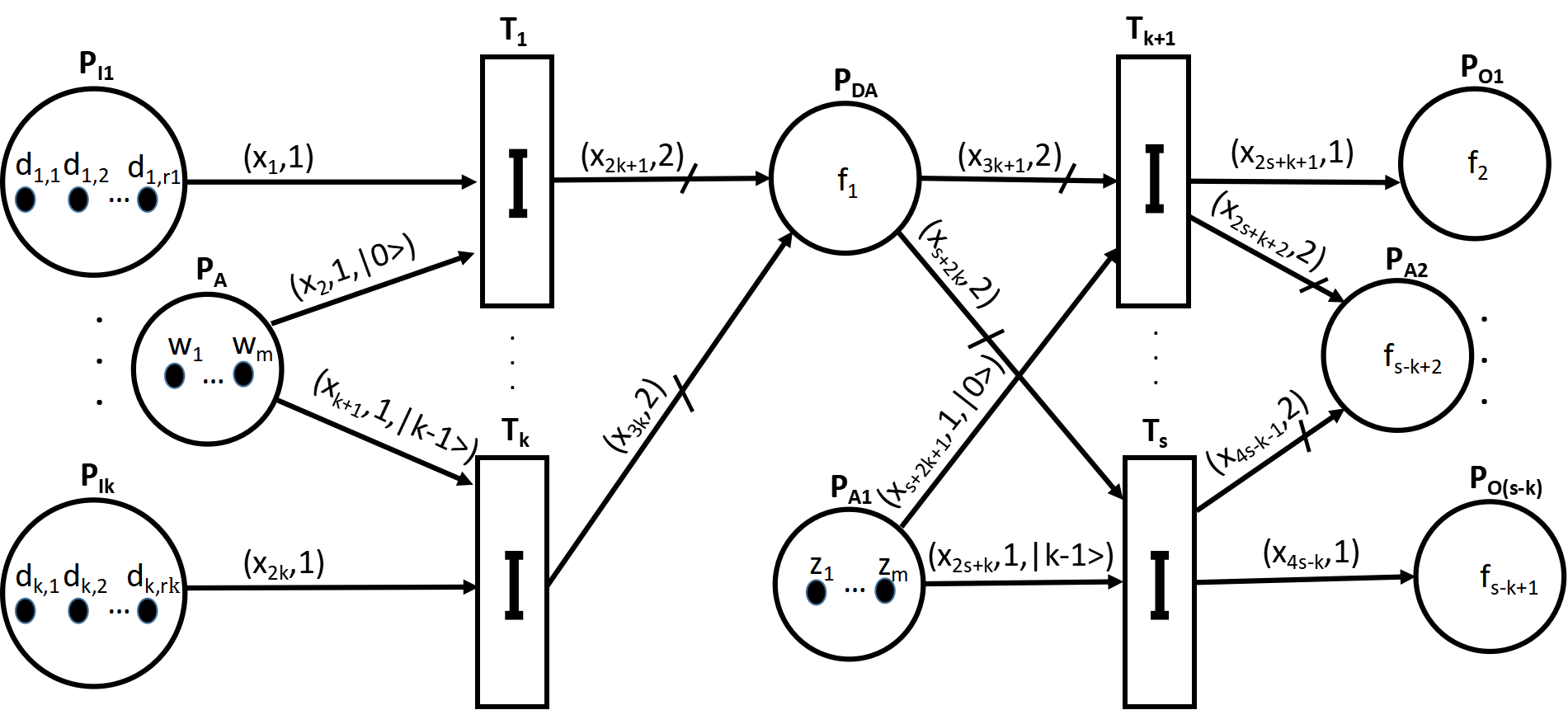}
\end{tabular}
\end{center}
\vspace{-10 pt}
\caption 
{ 
A multiple-input-multiple-output (MIMO) quantum buffer. } 
\label{fig:mimo}
\end{figure}

\begin{figure}
\vspace{-15pt}
\begin{center}
\begin{tabular}{c}
\includegraphics[height= 5.2 cm , width =15cm]{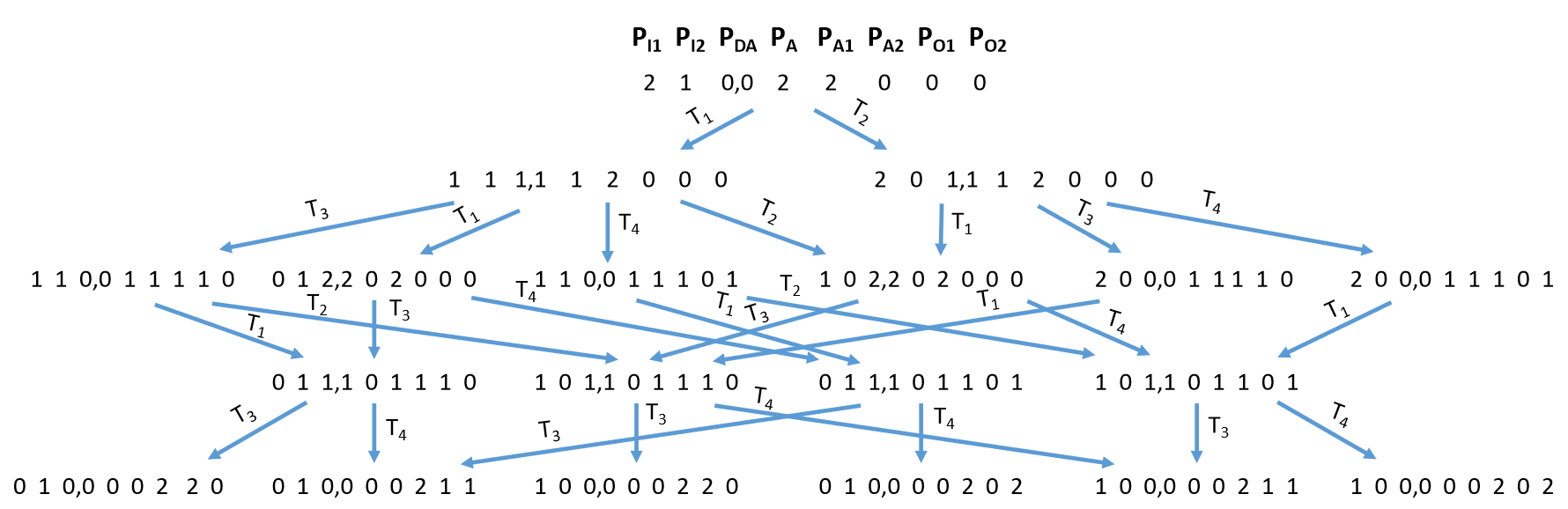}
\end{tabular}
\end{center}
\vspace{-10pt}
\caption 
{ 
The flow of $q$-tokens in a quantum MIMO buffer, $k = s  = 2, r_1 = 2, r_2 = 1, m = 2.$ }
\label{fig:mimoM}
\end{figure}

\vspace{-5pt}\noindent
A MIMO (Multiple Input, Multiple Output) quantum buffer is a type of quantum buffer that has multiple entry and multiple exit points for data $q$-tokens.  Two sets of ancillary $q$-tokens are used to select the input  and output places as the source and destination of a pair of transitions. The ancillary $q$-tokens $w_1,w_2,\cdots,w_m$ are used to select a $q$-token from an input place, and ancillary $q$-tokens $z_1,z_2,\cdots,z_m$ are used to select an output place to which the selected $q$-token is transferred.  Figure~\ref{fig:mimo} depicts the QPN model of MIMO quantum buffer of capacity $m,$ composed of $s$ transitions $T_1, T_2, T_3,..., T_{s}$, $k$ input places $P_{I1},P_{I2},...,P_{Ik}, s-k$ output places $P_{O1},P_{O2},...,P_{O(s-k)},$ two ancillary places $P_A,P_{A1},$ one joint data/ancillary place $P_{DA},$  $n = r_1 + r_2 +\cdots + r_k$ data $q$-tokens $d_{1,1}, d_{1,2}, ..., d_{(k-1),r(k-1)}, d_{k,rk}, $ and $2m$ ancillary $q$-tokens $w_1,w_2,\cdots\!,w_m, z_1,..., z_m$. 
The labels on the directed arcs $x_1, x_2,\cdots,x_{4s-k}$ denote the variables that represent $q$-tokens on the incoming and outgoing arcs of transitions. The data $q$-tokens are assigned to places $P_{I1},P_{I2},... P_{Ik},$ and $P_A$ as in MISO quantum buffer, and not repeated here. When the input transition $T_i$ fires, $1\le i\le k,$ it forwards the current data and ancillary $q$-token, say $d_{i,y}$ and $w_y$ to place $P_{DA}$ i.e., $f_1$ becomes $d_{i,y}, w_y$. When the output transition $T_j$  fires, $k+1\le j\le s,$ it forwards ancillary $q$-tokens $z_y$ and $w_y$ to $P_{A2},$ and data $q$-token $d_{i,y},$ to the corresponding output place $P_{oj}$. The behavior of the MIMO quantum buffer of capacity $m = 2$ with two input and two output places $P_{I1}, P_{I2}, P_{O1}$ and $P_{O2}$ is illustrated in Figure~\ref{fig:mimoM}. Input places $P_{I1}$  and $P_{I2}$  are initialized to 2 and 1 $q$-tokens, respectively. 
The quantities of $q$-tokens at the end of each  transition can be verified by tracing the $q$-tokens as they move between places. We note that at the end of possible transitions, the three $q$-tokens may be distributed from input places  $P_{I1}, P_{I2},$ to output places $P_{O1}$ and $P_{O2}$  in one of six ways: $(P_{I1}, P_{I2}, P_{O1}, P_{O2}) = (0,1,2,0), (0,1,1,1), (0,1,0,2), (1,0,2,0), (1,0,1,1), (1,0,0,2)$ as the diagram in Figure~\ref{fig:mimoM} shows. Note that some transition sequences  result in identical patterns of $q$-tokens in output places and are counted once. For example, both $T_1, T_3, T_1, T_4$ and $T_1, T_4,  T_1, T_3$  result in the placement of one token in each of $P_{o1}$ and $P_{o1}.$  It takes $2m$ transitions, to process $m$ data $q$-tokens. Overall, the whole process is reversible, and can be repeated again. 

\vspace{-5pt}
\subsection{Priority Quantum Buffer}
\vspace{-5pt}\noindent
A priority quantum buffer is a type of quantum buffer that prioritizes the processing of some $q$-tokens over others. The data   $q$-tokens are divided into two groups $P_{I1}, P_{I2},$ with those in $P_{I2}$ having a higher priority of transition over those in $P_{I1}.$ The priority is established using the ancillary $q$-tokens in $P_{A1}$ and $P_{A2}.$
\begin{figure}
\vspace{-45pt}
\begin{center}
\begin{tabular}{c}
\includegraphics[height= 7 cm,width=13 cm]{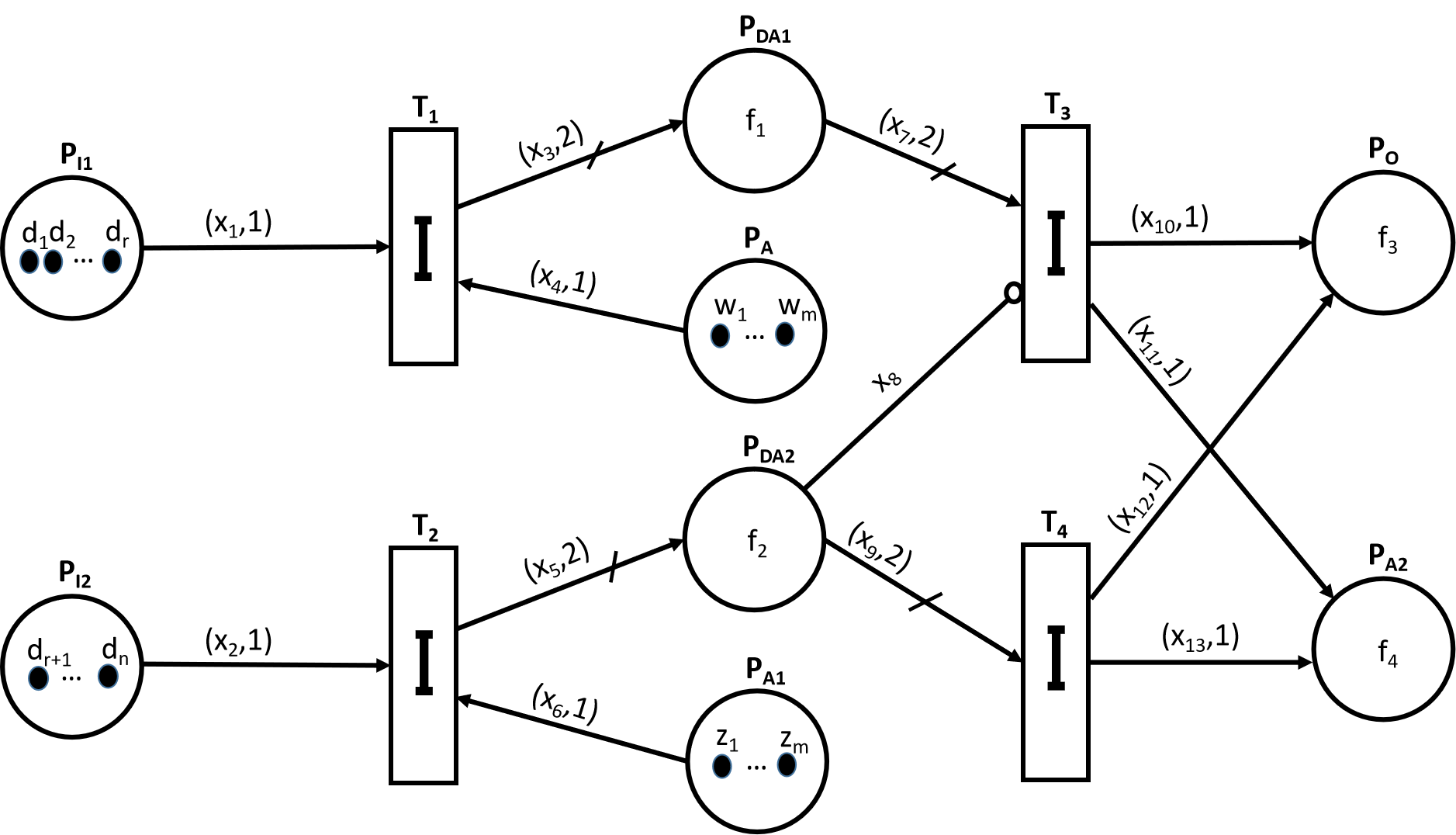}
\end{tabular}
\end{center}
\vspace{-5pt}
\caption 
{
A priority quantum buffer.} 
 \label{fig:priority}
\end{figure}
Figure~\ref{fig:priority} depicts the QPN model of a priority quantum buffer with capacity $m,$ four transitions $T_1, T_2, T_3, T_4$, eight places $P_{I1}, P_{I2}, P_A,  P_{A1}, P_{A2}, P_{DA1},  P_{DA2},  P_O$,  $r$ low priority data $q$-tokens, $d_1, d_2, ..., d_r$, $n-r$ high priority data $q$-tokens $d_{r+1},  d_{r+2}, ..., d_n$, $r+1\le j\le n,$  $m$ low priority ancillary $q$-tokens (qubits), $w_1,w_2,..., w_m$  and $m$ high priority ancillary $q$-tokens (qubits), $z_1,z_2,..., z_m$. The labels on the directed arcs $x_1, x_2,\cdots,x_{13}$ denote the variables that represent $q$-tokens on the incoming and outgoing arcs of transitions. It is assumed that the input transitions $T_1,T_2$ can run independently.   
The directed arc $x_8$ serves as an inhibitor arc to prevent the firing of output transition $T_3$ when there is any $q$-token in $P_{DA2}$. The $q$-tokens are assigned to places $P_{I1},P_{I2}, P_{A}$ and  $P_{A1}$ and are initialized  before the buffer starts operating as follows:

\vspace{-35pt}
\begin{align*}
 \mu(d_i) &= P_{I1}, 1\le i\le r,\\
 \mu(d_j) &= P_{I2}, r+1\le j\le n,\\
 \mu(w_u) &= P_{A}, 1\le u\le m,\\
 \mu(z_l) &= P_{A1}, 1\le l\le m
\end{align*}

\vspace{-30pt}
\begin{align*}
v(P_{I1},\mu, 0) = 
\left\{
\begin{array}{c}
d_1:d^*_1 \\
d_2:d^*_2 \\
 \cdot \\ [-5pt]
\cdot \\ [-5pt]
\cdot \\
d_r:d^*_r
\end{array} 
\right\}, 
v(P_{I2},\mu, 0) = 
\left\{
\begin{array}{c}
d_{r+1} :d^*_{r+1}\\
d_{r+2} :d^*_{r+2}\\
 \cdot \\ [-5pt]
\cdot \\ [-5pt]
\cdot \\
d_{n}:d^*_n
\end{array} 
\right\},
\end{align*}
\vspace{-20pt}
\begin{align*}
v(P_{A},\mu, 0) = 
\left\{
\begin{array}{c}
w_1 :|0\!>\\
w_2 :|0\!>\\
 \cdot \\ [-5pt]
\cdot \\ [-5pt]
\cdot \\
w_m:|0\!>
\end{array} 
\right\}, 
v(P_{A1},\mu, 0) = 
\left\{
\begin{array}{c}
z_1 :|0\!>\\
z_2 :|0\!>\\
 \cdot \\ [-5pt]
\cdot \\ [-5pt]
\cdot \\
z_m:|0\!>
\end{array} 
\right\}.
\end{align*}

$\,$\\

\begin{figure}
\vspace{-50pt}
\begin{center}
\begin{tabular}{c}
\includegraphics[height= 8.5 cm ,width = 14.5cm]{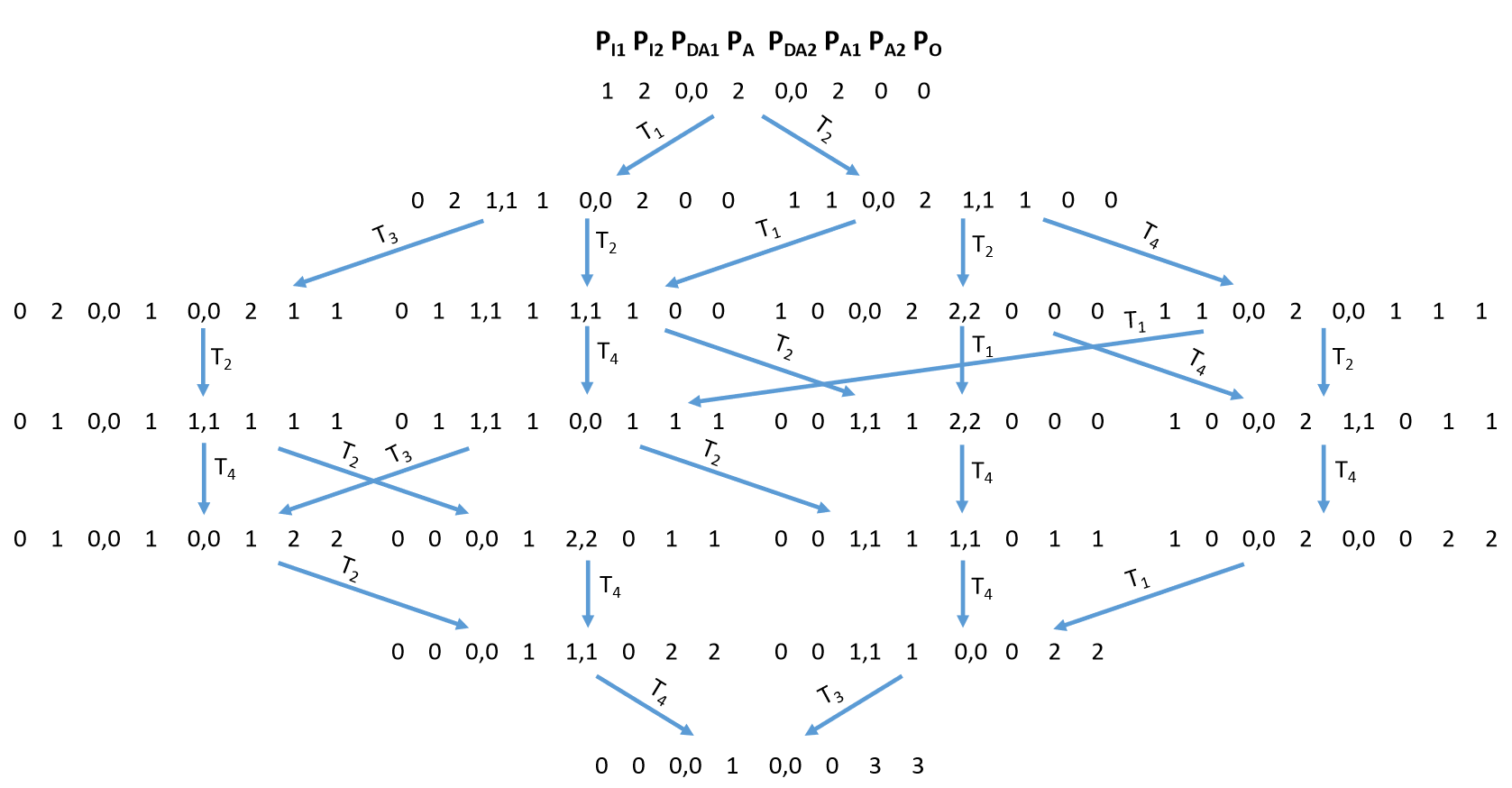}
\end{tabular}
\end{center}
\vspace{-5pt}
\caption 
{
The flow of $q$-tokens in a priority quantum buffer, $r = 1, n = 3, m = 3.$} 
 \label{fig:priorityM}
\end{figure}

\vspace{-40pt}\noindent
When the input transition, say $T_i$ fires, $1\le i\le 2,$ it forwards the current data and ancillary $q$-token, say $d_y$ and $z_y$ (or $w_y$) to place $P_{DAi}$ i.e., $f_i$ becomes either $d_y, z_y$ (or $d_y, w_y$). 
When the output transition, say $T_j$  fires, $3\le j\le 4,$ it forwards the  corresponding ancillary $q$-token $z_y$ or $w_y$ to $P_{A2}$ and data $q$-token $d_y$ to $P_O$.
However, the data $q$-tokens in place $P_{DA2}$ are prioritized over all tokens in place $P_{DA1}$ at the output place $P_O$ because transition $T_3$ cannot fire, due to the constraint of the inhibitor arc $x_8$. The behavior of the priority quantum buffer of capacity $m = 4$ is illustrated in Figure~\ref{fig:priorityM}. Input places $P_{I1}$  and $P_{I2}$  are initialized to one and two $q$-tokens, respectively. The quantities of $q$-tokens at the end of each  transition can be verified by tracing the $q$-tokens as they move between places. We note that some transition sequences inhibit the firing of transition $T_3$ until after $T_4$ is fired. For example, when $T_2$ fires, it inhibits the firing of transition $T_3$ unless $T_4$ fired.  It takes $2m$ transitions, to process $m$ data $q$-tokens. Overall, the whole process is reversible, and can be repeated again.

\vspace{-10pt}
\section{Quantum Runs and Simulations}
\label{runs}
\vspace{-8pt}

 In this section, we  present the quantum circuit implementations of the two Q-SR flip-flops we described earlier on the IBM quantum machine (IBM Brisbane, QPU-Eagle-R3, Version 1.1.41) using IBM Composer \cite{IBM-Composer}. We also validate the design of SISO, SIMO, MISO, MIMO, and priority quantum buffer on Jupyter \cite{Jupyter} using Python and IBM Qiskit \cite{IBM-Qiskit}. 

\vspace{-8pt}
\subsection*{A. Quantum S-R Flip-Flop Runs}
\vspace{-5pt}
 We implemented the two Q-S-R flip-flop designs on IBM Brisbane, QPU-Eagle-r3, Version 1.1.41 using 6 qubits.

 \vspace{-10pt}
\subsection*{A.1. Quantum S-R Flip-Flop Circuit 1 (Figure~\ref{fig:qsr1})}
 
\begin{figure}
\vspace{-40pt}
\begin{center}
\begin{tabular}{c}
\includegraphics[height= 2.8 cm,width=10cm]{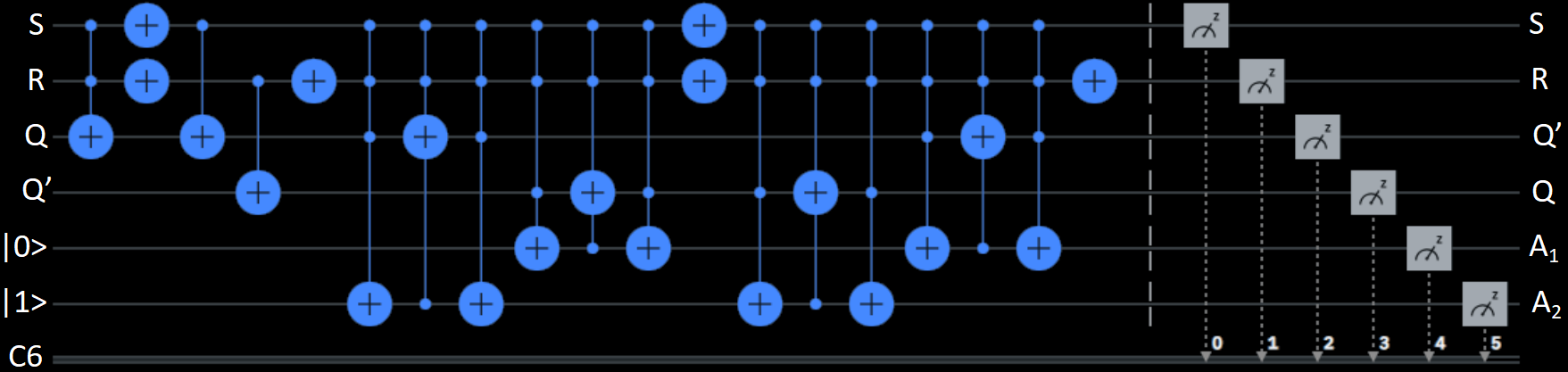}
\end{tabular}
\end{center}
\vspace{-10pt}
\caption 
{Quantum
Implementation of Q-S-R Flip-Flop Circuit 1. } 
 \label{fig:qsr-composer-1}
\end{figure}

\begin{figure}[!ht]
    \centering
    \begin{subfigure}[b]{0.4\textwidth}
        \centering
        \includegraphics[width=0.85\textwidth]{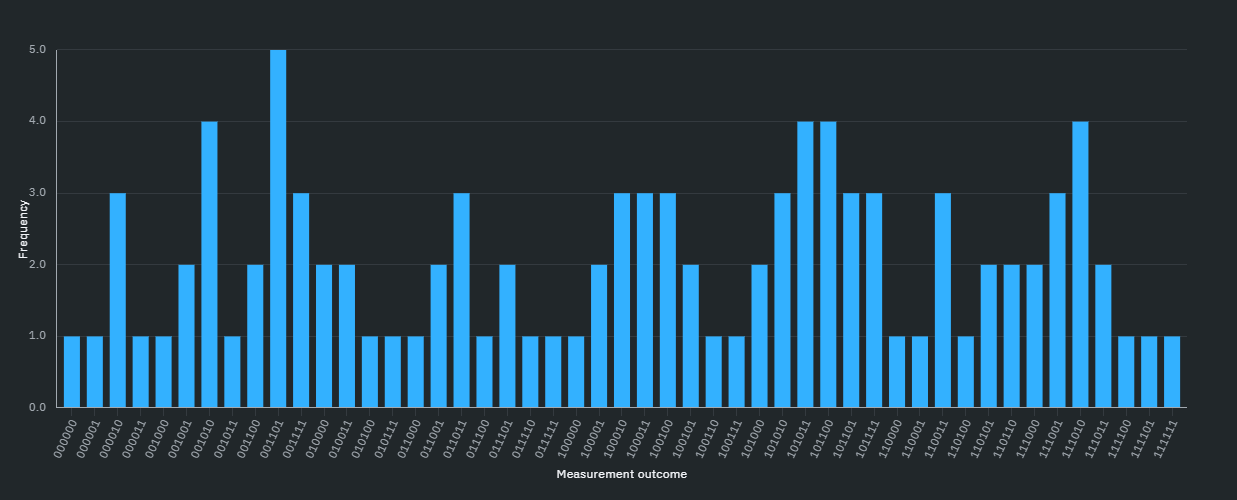}
        \caption{Initial input $SR = |10\rangle, QQ' = |10\rangle$}
        \label{fig:cs1q1}
    \end{subfigure}
    \hfill
    \begin{subfigure}[b]{0.4\textwidth}
        \centering
        \includegraphics[width=0.85\textwidth]{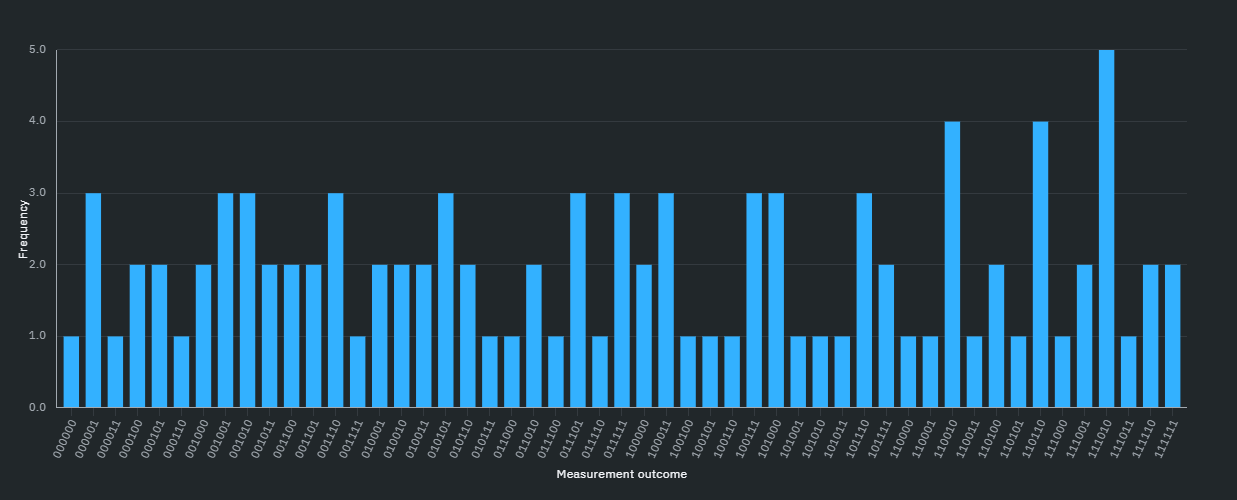}
        \caption{Initial input $SR = |10\rangle, QQ' = |01\rangle$}
        \label{fig:cs1qp1}
    \end{subfigure}
    \hfill
    \begin{subfigure}[b]{0.4\textwidth}
        \centering
        \includegraphics[width=0.85\textwidth]{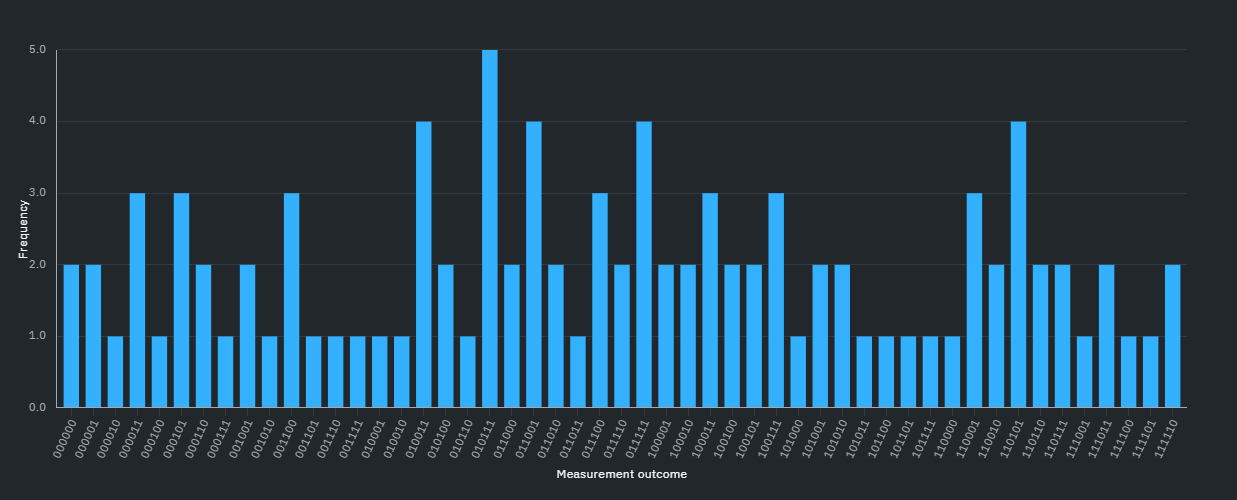}
        \caption{Initial input $SR = |01\rangle, QQ' = |10\rangle$}
        \label{fig:cr1q1}
    \end{subfigure}
      \hfill
    \begin{subfigure}[b]{0.4\textwidth}
        \centering
        \includegraphics[width=0.85\textwidth]{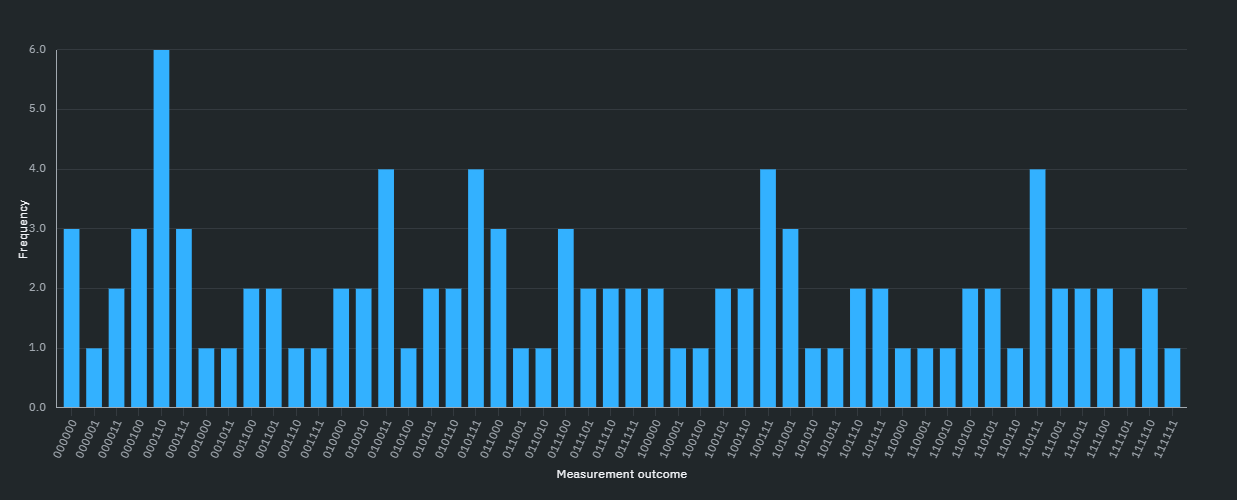}
        \caption{Initial input $SR = |01\rangle, QQ' = |01\rangle$}
        \label{fig:cr1qp1}
    \end{subfigure}
    \hfill
    \begin{subfigure}[b]{0.4\textwidth}
        \centering
        \includegraphics[width=0.85\textwidth]{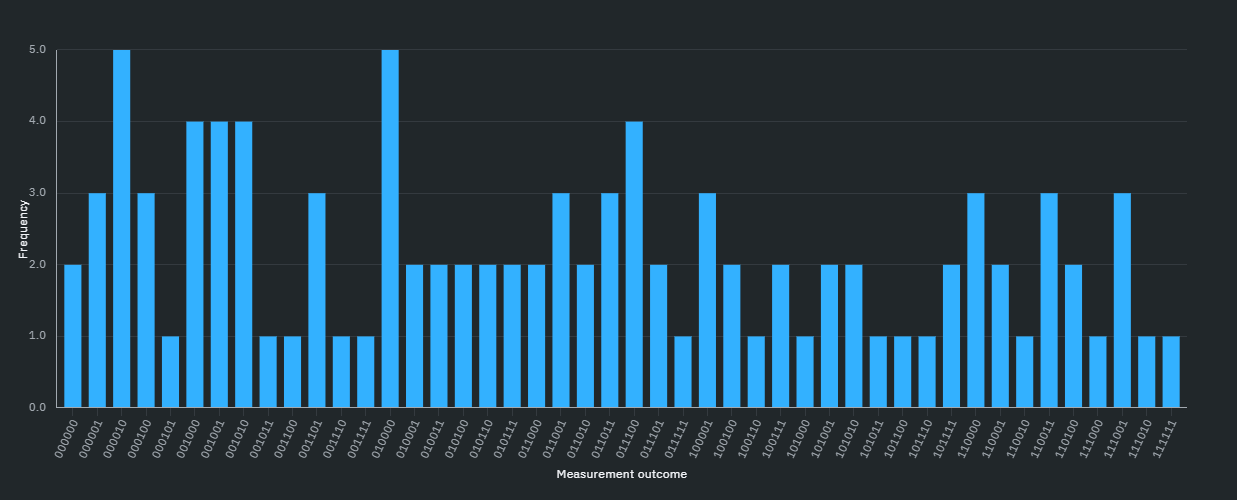}
        \caption{Initial input $SR = |00\rangle, QQ' = |10\rangle$}
        \label{fig:cs0r0q1}
    \end{subfigure}
    \hfill
    \begin{subfigure}[b]{0.4\textwidth}
        \centering
        \includegraphics[width=0.85\textwidth]{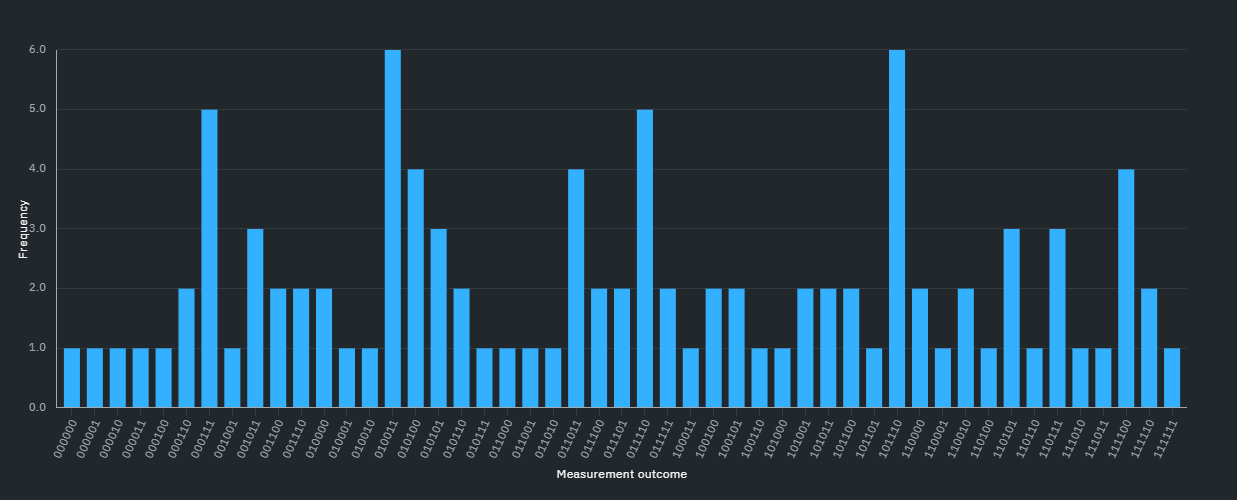}
        \caption{Initial input $SR = |00\rangle, QQ' = |01\rangle$}
        \label{fig:cs0r0qp1}
    \end{subfigure}
    
  \vspace{-5 pt}  
    \caption{Measurement outcomes of Q-S-R Flip-Flop Circuit $1.$}
    \label{fig:ibm-composer-result-1}
\end{figure}

 \vspace{-5pt}
 The quantum diagram for Circuit 1 was created on the IBM composer as shown in Figure~\ref{fig:qsr-composer-1}. 
  The circuit was tested under possible $SR$ configurations, ($|00\rangle$, $|10\rangle$, and $|01\rangle$), with the $QQ'$ state restricted to two valid present states, $|10\rangle$ and $|01\rangle$. Each test case was run 100 times, and the results are shown in Figures~\ref{fig:ibm-composer-result-1}. The vertical axis represents the frequency of the output occurrences, while the  horizontal axis corresponds to the sequence of output states: $SRQ'QA_1A_2,$ where $A_1$ and $A_2$ represent ancillary qubits. We analyzed both the combined and individual measurement  outcomes such as SRQQ$'$, QQ$'$, and Q, to ensure a thorough evaluation of the circuit's behavior under different input configurations as shown in Table~\ref{tab:circuit-1}. For the set input, i.e., $|SR\rangle = |10\rangle,$ the Q-S-R flip-flop's combined and individual next states, $|SRQQ'\rangle= |1010\rangle , |QQ'\rangle=|10\rangle$ and $|Q\rangle=|1\rangle$ were measured   $7,34,62$ and $6,28,53$ times, respectively in the present states $|QQ'\rangle=|10\rangle$  and  $|QQ'\rangle=|01\rangle$. For the reset input, i.e.,  $|SR\rangle = |01\rangle,$ the combined and individual next states, $|SRQQ'\rangle= |0101\rangle , |QQ'\rangle=|01\rangle$ and $|Q=0\rangle$  were measured $5,30,57$ and $11,39,59$ times, respectively in the present states $|QQ'\rangle=|10\rangle$  and  $|QQ'\rangle=|01\rangle$. When $|SR\rangle$ is set to $|00\rangle$, the combined and individual next states, $|SRQQ'\rangle= |1010\rangle , |QQ'\rangle=|10\rangle$ and $|Q=1\rangle$ were measured   $8,34,52$ times in the present state, $|QQ'\rangle=|10\rangle$ and 
  $|SRQQ'\rangle= |0101\rangle , QQ'\rangle=|01\rangle$ and $|Q=0\rangle$ were measured $8,31,51$ in the present state $|QQ'\rangle=|01\rangle$.
  
\vspace{15pt}

\begin{table}[h!]
\vspace{15pt}
{\footnotesize
    \centering
    \begin{tabular}{|c|c|c|c?c|c|c|}
        \hline
        \multicolumn{4}{|c?}{\textbf{Input}} & \multicolumn{3}{c|}{\textbf{ Q Output Counts}} \\
        \hline
        S & R & Q & Q$'$ & SRQQ$'$ & QQ$'$ & Q \\
        \hline
        $|1\rangle$ & $|0\rangle$ & $|1\rangle$ & $|0\rangle$ & $|1010\rangle,  7$ & $|10\rangle, 34$ & $|1\rangle, 62$  \\
        \hline
        $|1\rangle$ & $|0\rangle$ & $|0\rangle$ & $|1\rangle$ & $|1010\rangle,  6$ & $|10\rangle, 28$ & $|1\rangle, 53$ \\
        \hline
        $|0\rangle$ & $|1\rangle$ & $|1\rangle$ & $|0\rangle$ & $|0101\rangle,  5$ & $|01\rangle, 30$ & $|0\rangle, 57$ \\
        \hline
        $|0\rangle$ & $|1\rangle$ & $|0\rangle$ & $|1\rangle$ & $\,\,\,|0101\rangle, 11$ & $|01\rangle, 39$ & $|0\rangle, 59$ \\
        \hline
         $|0\rangle$ & $|0\rangle$ & $|1\rangle$ & $|0\rangle$ & $|0010\rangle, 8$ & $|10\rangle, 34$ & $|1\rangle, 52$ \\
        \hline
        $|0\rangle$ & $|0\rangle$ & $|0\rangle$ & $|1\rangle$ & $|0001\rangle,  8$ & $|01\rangle, 31$ & $|0\rangle, 51$ \\
        \hline
    \end{tabular}
    \caption{Compiled measurement outcomes of Q-S-R Flip-Flop Circuit $1.$}
    \label{tab:circuit-1}
    }
\end{table}

 $\,$\\
 \vspace{-40pt}
\subsection*{A.2. Quantum S-R Flip-Flop Circuit 2 (Figure~\ref{fig:qsr2})}
 
\begin{figure}[t!]
\vspace{-20pt}
\centering
\begin{center}
\begin{tabular}{c}
\includegraphics[height= 2.5 cm,width=7.5cm]{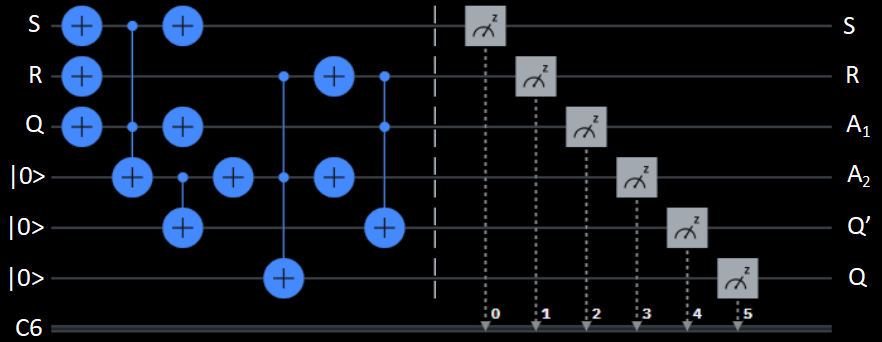}
\end{tabular}
\end{center}
\vspace{-10pt}
\caption 
{Circuit $2$ implementation of a Q-S-R flip-flop. }
 \label{fig:qsr-composer-2}
\end{figure}

\begin{figure}[t!]
\vspace{-1pt}
    \centering
    \begin{subfigure}[b]{0.4\textwidth}
        \centering
        \includegraphics[width=0.85\textwidth]{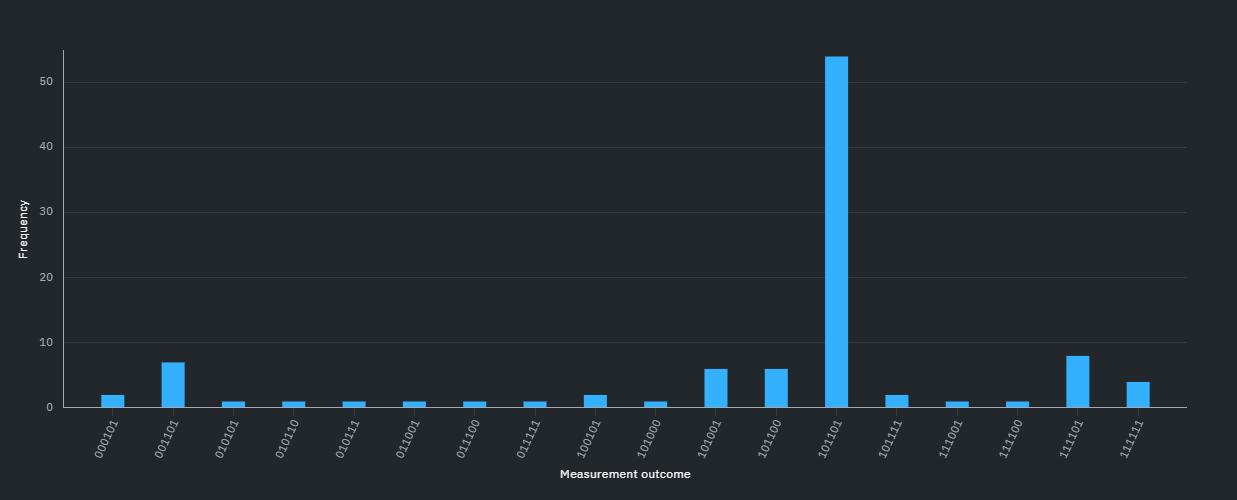}
        \caption{Initial input $SR = |10\rangle, Q = |1\rangle$}
        \label{fig:c2s1q1}
    \end{subfigure}
    \hfill
    \begin{subfigure}[b]{0.4\textwidth}
        \centering
        \includegraphics[width=0.85\textwidth]{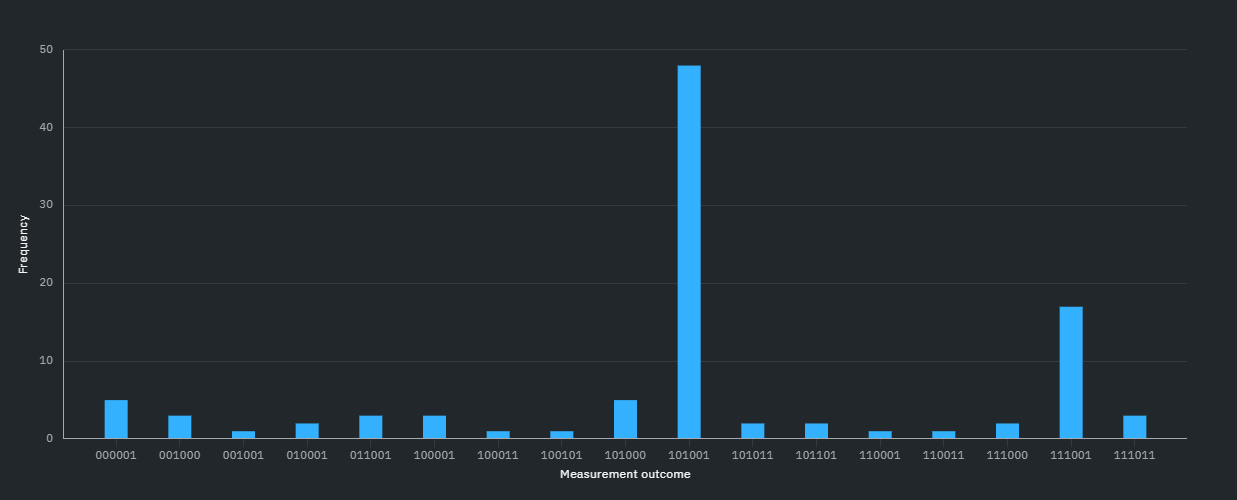}
        \caption{Intial input $SR = |10\rangle, Q = |0\rangle$}
        \label{fig:c2s1q0}
    \end{subfigure}
    \hfill
    \begin{subfigure}[b]{0.4\textwidth}
        \centering
        \includegraphics[width=0.85\textwidth]{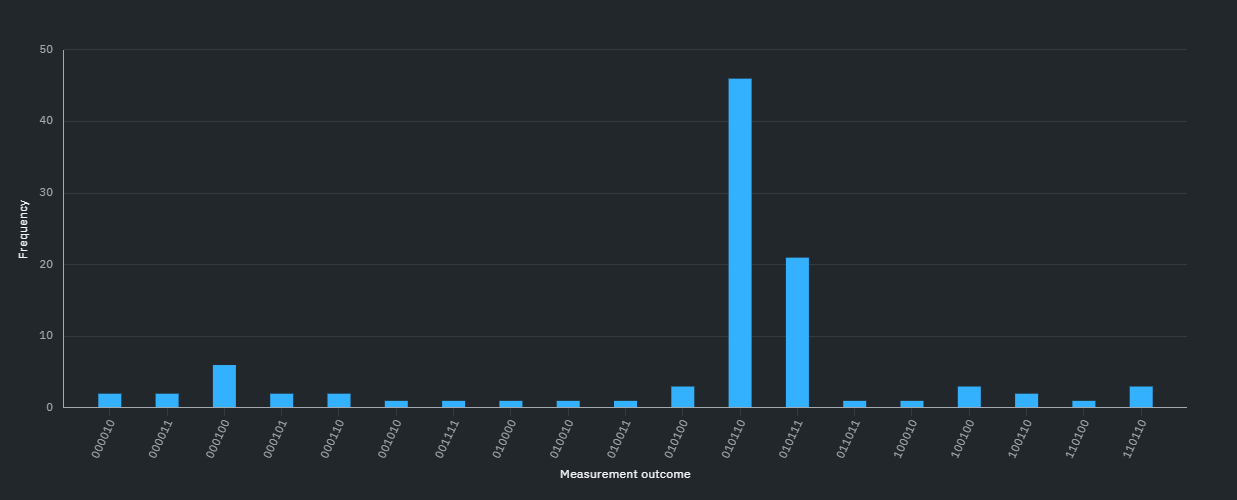}
        \caption{Initial input $SR = |01\rangle, Q = |1\rangle$}
        \label{fig:c2r1q1}
    \end{subfigure}
      \hfill
    \begin{subfigure}[b]{0.4\textwidth}
        \centering
        \includegraphics[width=0.85\textwidth]{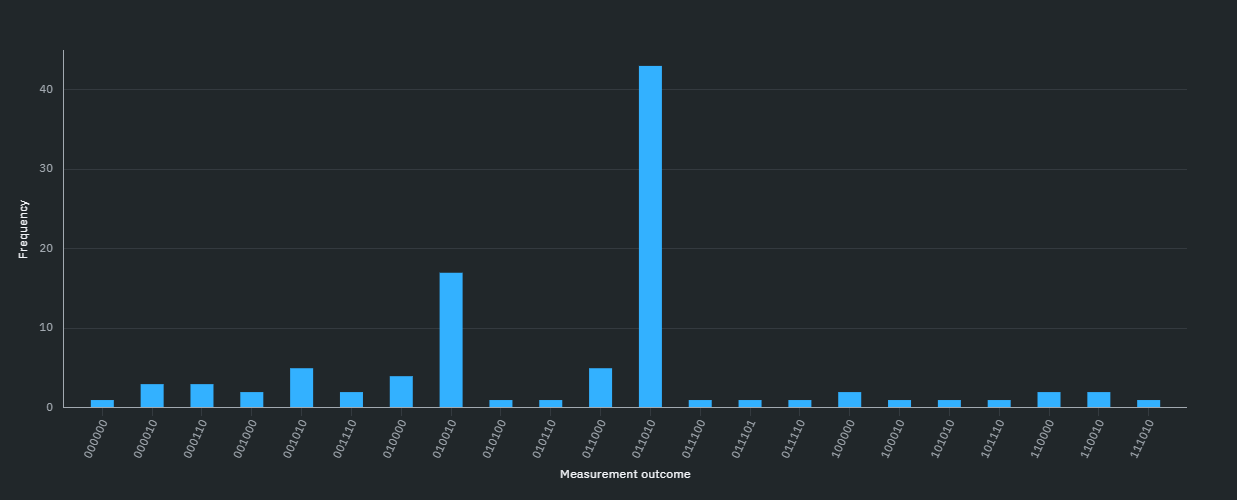}
        \caption{Initial input $SR = |01\rangle, Q = |0\rangle$}
        \label{fig:c2r1q0}
    \end{subfigure}
    \hfill
    \begin{subfigure}[b]{0.40\textwidth}
        \centering
        \includegraphics[width=0.85\textwidth]{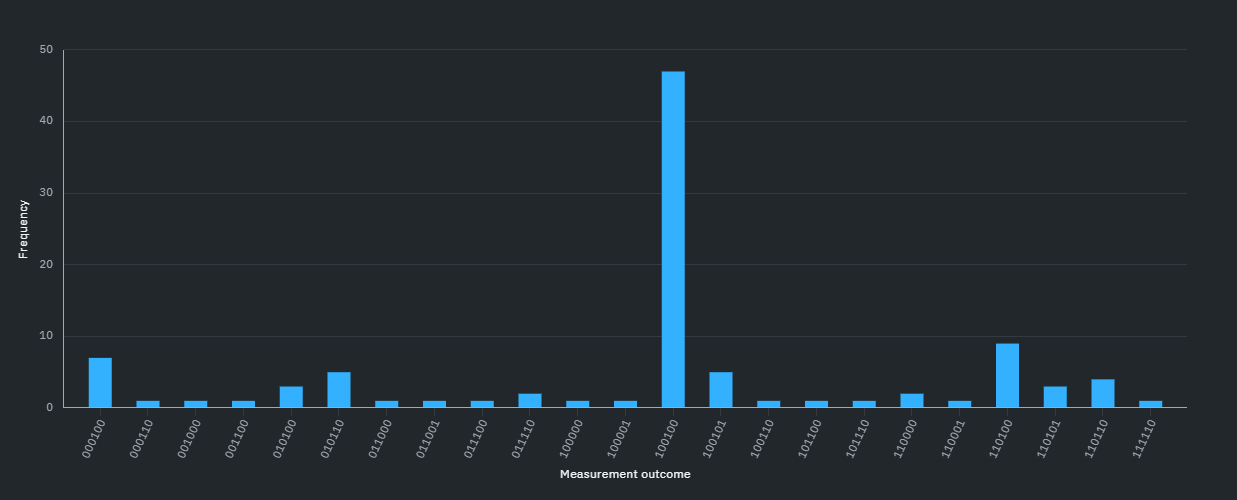}
        \caption{Initial input $SR = |00\rangle, Q = |1\rangle$}
        \label{fig:c2s0r0q1}
    \end{subfigure}
    \hfill
    \begin{subfigure}[b]{0.40\textwidth}
        \centering
        \includegraphics[width=0.85\textwidth]{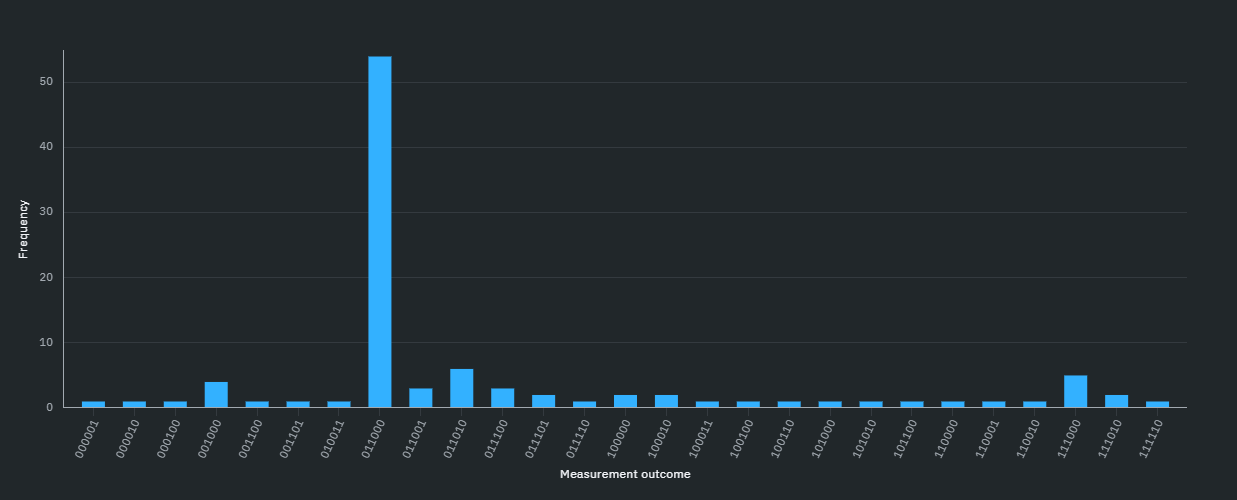}
        \caption{Initial input $SR = |00\rangle, Q = |0\rangle$}
        \label{fig:c2s0r0q0}
    \end{subfigure}
    
  \vspace{1 pt}  
    \caption{Circuit 2 measurement outcomes.}
    \label{fig:ibm-composer-result-2}
\end{figure}

 \vspace{-3pt}
 The quantum Circuit 2 was created on the IBM composer as shown in Figure~\ref{fig:qsr-composer-2}. 
  Like Circuit 1,  Circuit 2 was also tested under possible $SR$ configurations, ($|00\rangle$, $|10\rangle$, and $|01\rangle$) with the $QQ'$ state restricted to two valid present states, $|10\rangle$ and $|01\rangle$. Each test case was run 100 times, and the results are shown in Figures~\ref{fig:ibm-composer-result-2}. The vertical axis represents the frequency of the output occurrences, while the horizontal axis corresponds to the sequence of output states: $SRA_1A_2Q'Q,$  where $A_1$ and $A_2$ denote ancillary qubits as before. The circuit's behavior under different input configurations is tabulated for different sets of output measurements in Table~\ref{tab:circuit-2}. For the set input, i.e., $|SR\rangle = |10\rangle,$ the Q-S-R flip-flop's combined and individual next states, $|SRQQ'\rangle= |1010\rangle , |QQ'\rangle=|10\rangle$ and $|Q\rangle=|1\rangle$ were measured   $63,71,85$ and $53,62,86$ times, respectively in the present states $|QQ'\rangle=|10\rangle$  and  $|QQ'\rangle=|01\rangle$. For the reset input, i.e.,  $|SR\rangle = |01\rangle,$ the combined and individual next states, $|SRQQ'\rangle= |0101\rangle , |QQ'\rangle=|01\rangle$ and $|Q=0\rangle$  were measured $47,74,90$ and $63,75,90$ times, respectively in the present states $|QQ'\rangle=|10\rangle$  and  $|QQ'\rangle=|01\rangle$. When $|SR\rangle$ is set to $|00\rangle$, the combined and individual next states, $|SRQQ'\rangle= |1010\rangle , |QQ'\rangle=|10\rangle$ and $|Q=1\rangle$ were measured   $49,57,77$ times in the present states $|QQ'\rangle=|10\rangle$ and 
  $|SRQQ'\rangle= |0101\rangle , QQ'\rangle=|01\rangle$ and $|Q=0\rangle$ were measured $57,71,79$ in the present states $|QQ'\rangle=|01\rangle$.

\vspace{20pt}
\begin{table}[t!]
    {\footnotesize \centering
    \begin{tabular}{|c|c|c?c|c|c|}
        \hline
        \multicolumn{3}{|c?}{\textbf{Input}} & \multicolumn{3}{c|}{\textbf{Output}} \\
        \hline
        S & R & Q  & SRQQ$'$ & QQ$'$ & Q \\
        \hline
        $|1\rangle$ & $|0\rangle$ & $|1\rangle$ & $|1010\rangle = 63$ & $|10\rangle = 71$ & $|1\rangle = 85$  \\
        \hline
        $|1\rangle$ & $|0\rangle$ & $|0\rangle$ & $|1010\rangle = 53$ & $|10\rangle = 62$ & $|1\rangle = 86$ \\
        \hline
        $|0\rangle$ & $|1\rangle$ & $|1\rangle$ & $|0101\rangle = 47$ & $|01\rangle = 74$ & $|0\rangle = 90$ \\
        \hline
        $|0\rangle$ & $|1\rangle$ & $|0\rangle$ & $|0101\rangle = 63$ & $|01\rangle = 75$ & $|0\rangle = 90$ \\
        \hline
         $|0\rangle$ & $|0\rangle$ & $|1\rangle$ & $|0010\rangle = 49$ & $|10\rangle = 57$ & $|1\rangle = 77$ \\
        \hline
        $|0\rangle$ & $|0\rangle$ & $|0\rangle$ & $|0001\rangle = 57$ & $|01\rangle = 71$ & $|0\rangle = 79$ \\
        \hline
    \end{tabular}
    \caption{Compiled measurement outcomes of Q-S-R flip-flop Circuit $2.$}
    \label{tab:circuit-2}
    }
\end{table}

\subsection*{A.3. Quantum S-R  Flip-Flop Via A Quantum J-K Flip-Flop}

\begin{figure}[t!]
\vspace{-15pt}
\centering
\begin{center}
\begin{tabular}{c}
\includegraphics[height=2.8 cm,width=9cm]{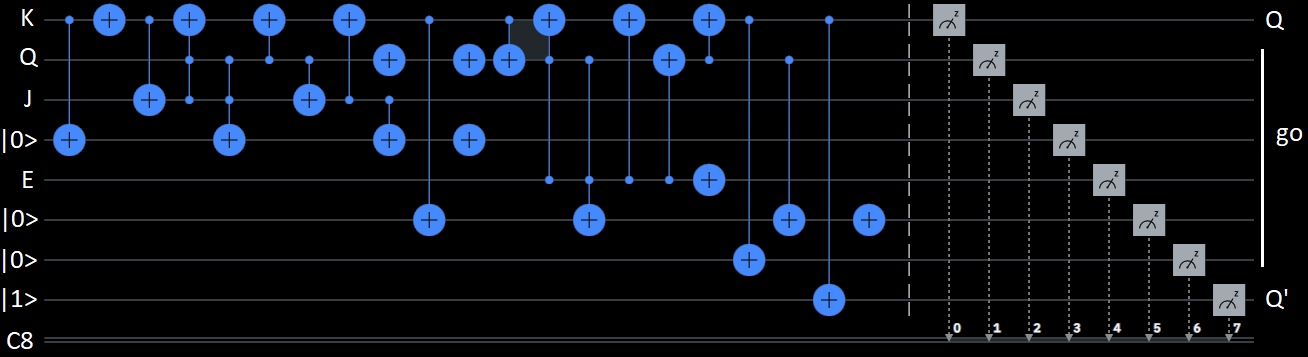}
\end{tabular}
\end{center}
\vspace{-10pt}
\caption 
{Implementation of a Q-J-K flip-flop circuit~\cite{misra-2024}.}  
\label{fig:qsr-composer-jk}
\end{figure}

\begin{figure}[t!]
\vspace{1pt}
    \centering
    \begin{subfigure}[b]{0.4\textwidth}
        \centering
        \includegraphics[width=0.85\textwidth]{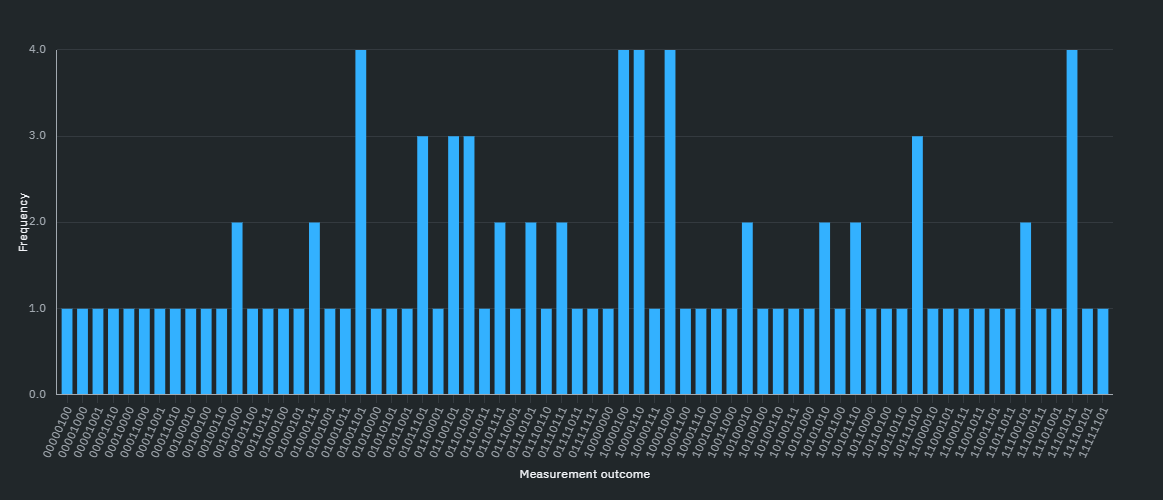}
        \caption{Initial input $JK = |10\rangle, Q = |1\rangle$}
        \label{fig:j1q1}
    \end{subfigure}
    \hfill
    \begin{subfigure}[b]{0.4\textwidth}
        \centering
        \includegraphics[width=0.85\textwidth]{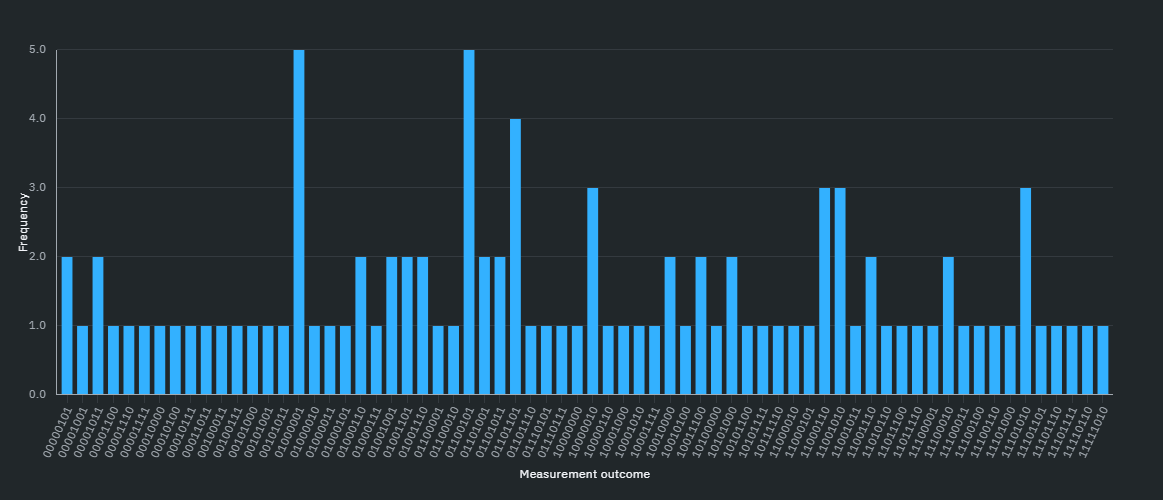}
        \caption{Intial input $JK = |10\rangle, Q = |0\rangle$}
        \label{fig:j1q0}
    \end{subfigure}
    \hfill
    \begin{subfigure}[b]{0.4\textwidth}
        \centering
        \includegraphics[width=0.85\textwidth]{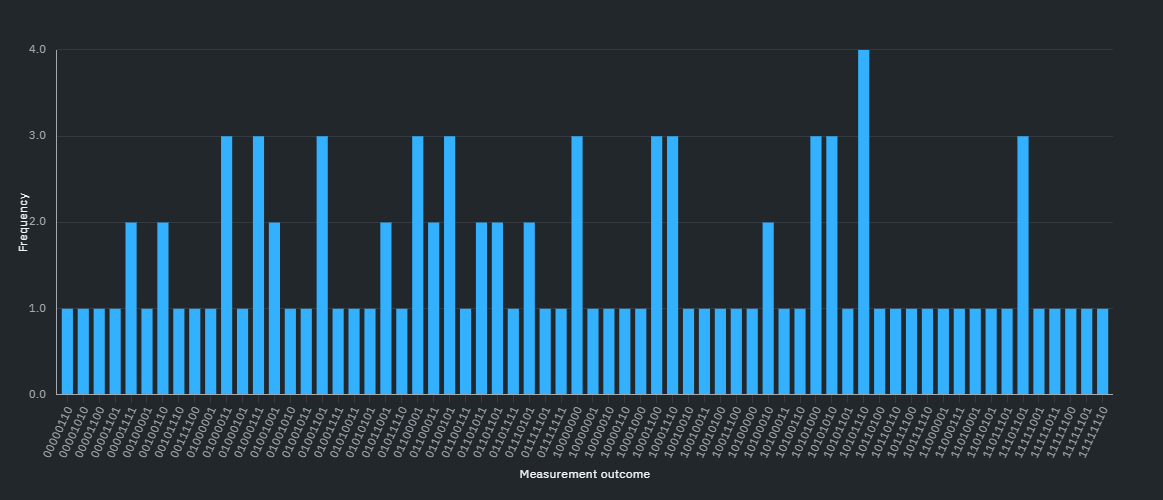}
        \caption{Initial input $JK = |01\rangle, Q = |1\rangle$}
        \label{fig:k1q1}
    \end{subfigure}
      \hfill
    \begin{subfigure}[b]{0.4\textwidth}
        \centering
        \includegraphics[width=0.85\textwidth]{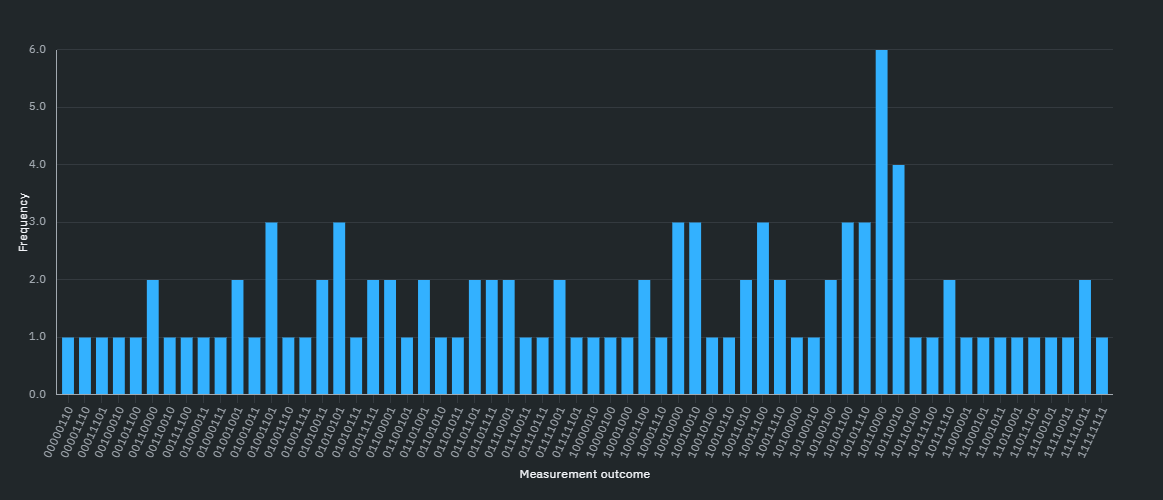}
        \caption{Initial input $JK = |01\rangle, Q = |0\rangle$}
        \label{fig:k1q0}
    \end{subfigure}
    \hfill
    \begin{subfigure}[b]{0.40\textwidth}
        \centering
        \includegraphics[width=0.85\textwidth]{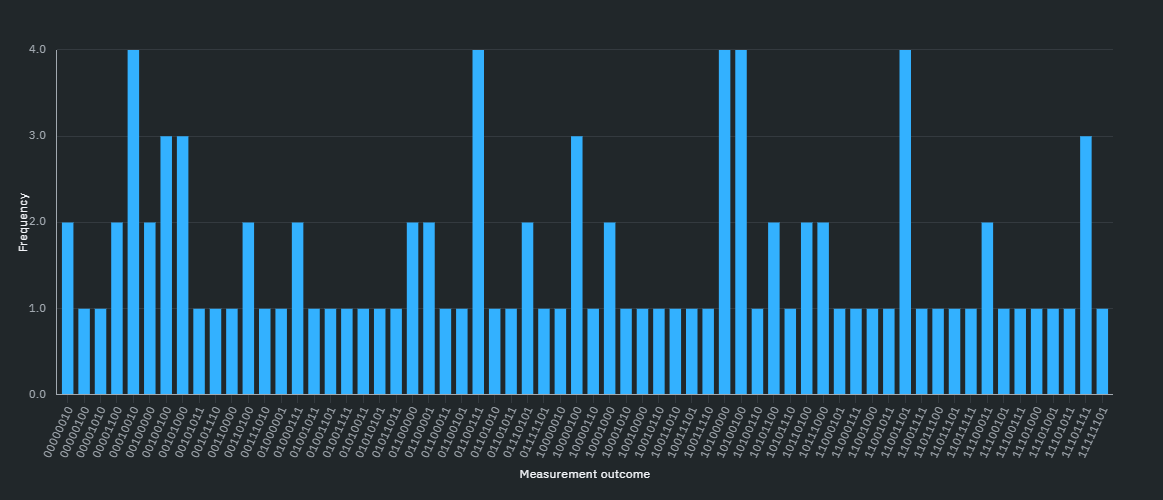}
        \caption{Initial input $JK = |00\rangle, Q = |1\rangle$}
        \label{fig:j0k0q1}
    \end{subfigure}
    \hfill
    \begin{subfigure}[b]{0.40\textwidth}
        \centering
        \includegraphics[width=0.85\textwidth]{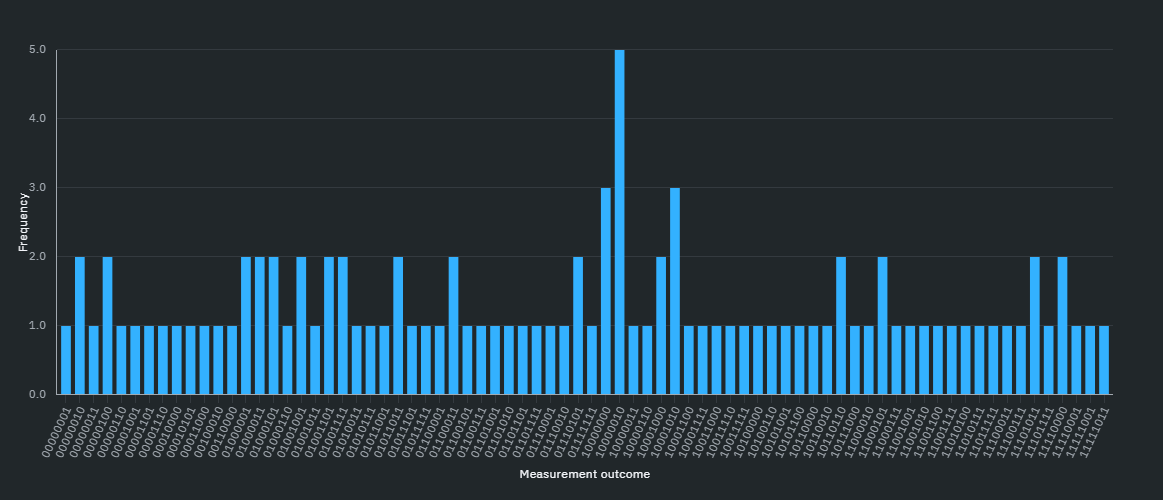}
        \caption{Initial input $JK = |00\rangle, Q = |0\rangle$}
        \label{fig:j0k0q0}
    \end{subfigure}
    
  \vspace{1 pt}  
    \caption{Q-J-K circuit measurement outcomes.}
    \label{fig:ibm-composer-JK-result-2}
\end{figure}

\vspace{-5pt}
As stated in Section~\ref{quantumFlip-Flops}, a quantum S-R flip-flop can be obtained from a quantum J-K flip-flop with  $(J = |0\rangle, K = |0\rangle), (J = |0\rangle, K = |1\rangle), (J = |1\rangle, K = |0\rangle)$ settings, and avoiding the $(J = |1\rangle, K = |1\rangle)$ setting. To compare the performance of the two Q-S-R flip-flop circuits to the performance of  a quantum J-K flip-flop, we implemented the quantum J-K flip-flop that was reported in~\cite{misra-2024} on the IBM Brisbane, QPU-Eagle-R3, Version 1.1.41 as shown in Figure~\ref{fig:qsr-composer-jk}.  
\begin{table}[t!]
{\footnotesize
   \centering
   \begin{tabular}{|c|c|c?c|c|}
       \hline
       \multicolumn{3}{|c?}{\textbf{Input}} & \multicolumn{2}{c|}{\textbf{Output}} \\
       \hline
       J & K & Q  & QQ$'$ & Q \\
       \hline
       $|1\rangle$ & $|0\rangle$ & $|1\rangle$ & $|10\rangle = 34$ & $|1\rangle = 51$  \\
       \hline
       $|1\rangle$ & $|0\rangle$ & $|0\rangle$ & $|10\rangle = 42$ & $|1\rangle = 50$ \\
       \hline
       $|0\rangle$ & $|1\rangle$ & $|1\rangle$ & $|01\rangle = 35$ & $|0\rangle = 44$ \\
       \hline
       $|0\rangle$ & $|1\rangle$ & $|0\rangle$ & $|01\rangle = 45$ & $|0\rangle = 56$ \\
       \hline
        $|0\rangle$ & $|0\rangle$ & $|1\rangle$  & $|10\rangle = 22$ & $|1\rangle = 43$ \\
       \hline
       $|0\rangle$ & $|0\rangle$ & $|0\rangle$ & $|01\rangle = 31$ & $|0\rangle = 44$ \\
       \hline
   \end{tabular}
   \caption{Compiled measurement outcomes of Q-J-K Flip-Flop.}
   \label{tab:circuit-3}
   }
\end{table}
Table~\ref{tab:circuit-3} summarizes the measurements for $QQ'$ and $Q$ outputs for possible $JK$ configurations, ($|00\rangle$, $|10\rangle$, and $|01\rangle$)  with the $QQ'$ state restricted to two valid present states, $|10\rangle$ and $|01\rangle.$ The results indicate that, in terms of valid next states measurements,  Circuit 2 implementation of the Q-S-R flip-flop outperforms both Circuit 1 implementation of the Q-S-R flip-flop and the implementation of Q-J-K flip-flop in~\cite{misra-2024} by a significant margin while Circuit 1 implementation of the Q-S-R flip-flop and the implementation of Q-J-K flip-flop have comparable correct measurements of  $QQ'$  and $Q$ states. We also tabulated the quantum gate counts of the three quantum flip-flop implementations as well as the Q-J-K flip-flop design that was described in~\cite{misra-2018} in Table~\ref{tab:costFlipFlop}. The implementation of this Q-J-K flip-flop was not carried out as it uses a CV$^{+}$ gate that is not directly available in IBM Composer and OpenQasm 2.0.  Overall, Circuit 2 implementation of S-R-Q flip-flop outperforms the other  quantum flip-flop implementations in terms of the validity of its output states and cost.

\begin{table}[h!] 
\vspace{10pt}
 {\footnotesize   \centering
    \begin{tabular}{|c?c|c|c|c|c|c|} 
    \hline
    &NOT & CV & CV$^{+}$ & C-NOT &CC-NOT &CCC-NOT \\ 
    \hline
    Quantum J-K~\cite{misra-2018} & $-$ & $2$ &$1$ & $8$ & $-$ & $-$\\ 
    \hline
    Quantum J-K~\cite{misra-2024} & $6$ & $-$ &$-$ & $14$ & $4$ &$-$
    \\ 
    \hline
    Circuit-1 & $6$ & $-$ &$-$ & $2$ & $1$ &$12$\\ 
    \hline
    Circuit-2 & $8$ & $-$ &$-$ & $1$ & $3$ &$-$\\ 
   \hline
 \end{tabular}

 \vspace{1pt}
    \caption{Gate counts of Q-S-R and Q-J-K flip-flops.}
    \label{tab:costFlipFlop} 
}\end{table}

\vspace{10pt}
\subsection*{B. Quantum Buffer Simulations}
\vspace{-5pt}
 We simulated all quantum buffer designs on Jupyter \cite{Jupyter} in Python and IBM Qiskit \cite{IBM-Qiskit}. 
 
 \vspace{-10pt}
\subsection*{B.1. The SISO QPN Simulations}

\begin{figure}[!ht]
\vspace{-15pt}
    \centering
    \begin{subfigure}[b]{0.60\textwidth}
        \centering \includegraphics[width=\textwidth]{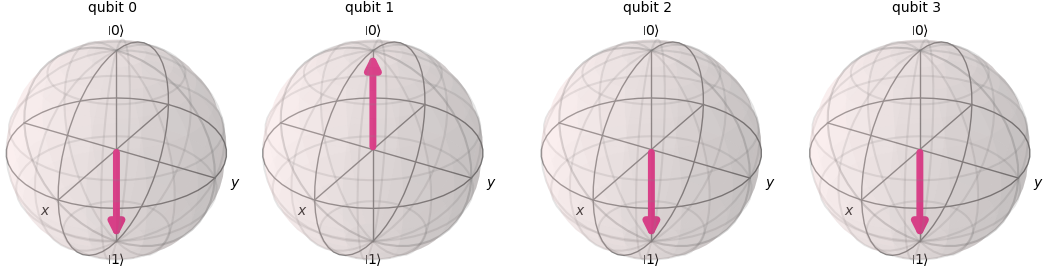}
        \caption{$d_1, d_2, d_3$ in $P_I$}
        \label{fig:d1011}
    \end{subfigure}
    \hfill
    \begin{subfigure}[b]{0.30\textwidth}
        \centering \includegraphics[width=\textwidth]{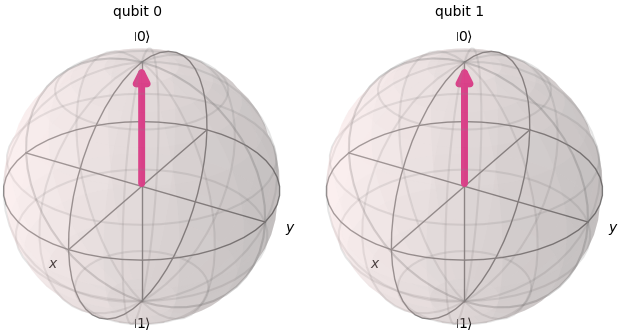}
        \caption{$z_1, z_2$ in $P_A$}
        \label{fig:z00}
    \end{subfigure}
    \hfill
    \begin{subfigure}[b]{0.30\textwidth}
        \centering
        \includegraphics[width=\textwidth]{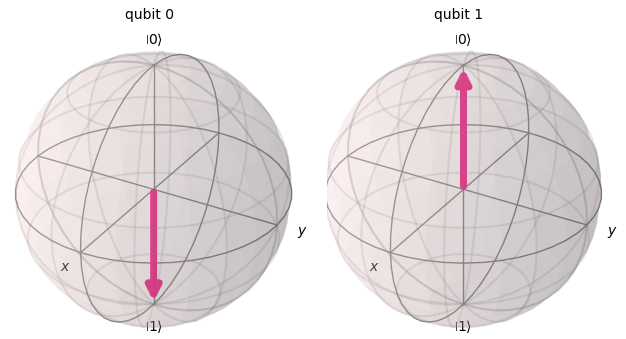}
        \caption{$d_1$ in $P_{O}$}
        \label{fig:d10}
    \end{subfigure}
    \hfill
    \begin{subfigure}[b]{0.15\textwidth}
        \centering
        \includegraphics[width=\textwidth]{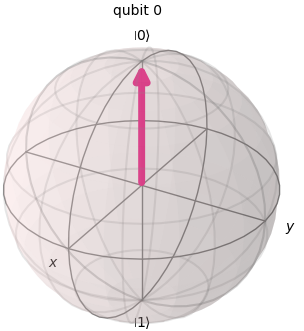}
        \caption{$z_1$ in $P_{A1}$}
        \label{fig:z0_pa1}
    \end{subfigure}
    \hfill
    \begin{subfigure}[b]{0.15\textwidth}
        \centering
        \includegraphics[width=\textwidth]{images/z0.png}
        \caption{$z_2$ in $P_{A}$}
        \label{fig:z0_pa}
    \end{subfigure}
    \hfill
    \begin{subfigure}[b]{0.30\textwidth}
        \centering    \includegraphics[width=\textwidth]{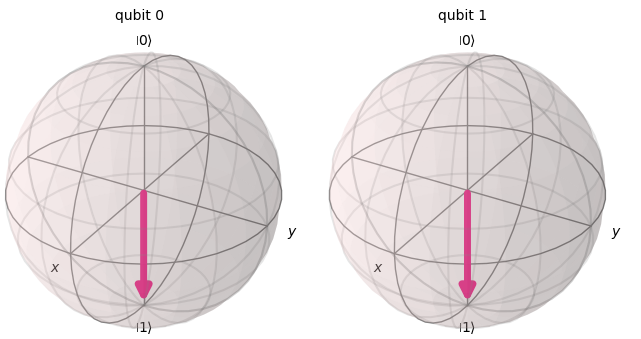}
        \caption{$d_2, d_3$ in $P_I$}
        \label{fig:d11}
    \end{subfigure}
    \hfill
    
    \begin{subfigure}[b]{0.45\textwidth}
        \centering
        \includegraphics[width=\textwidth]{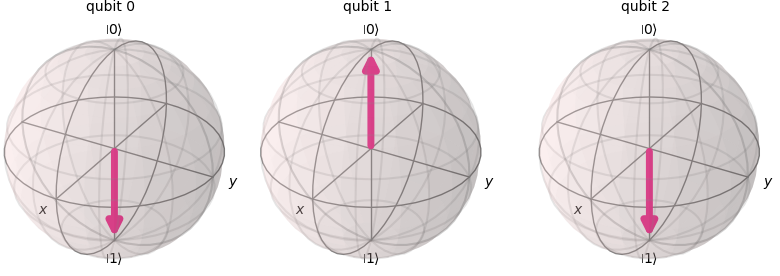}
        \caption{$d_1, d_2$ in $P_O$}
        \label{fig:d101}
    \end{subfigure}
    \hfill
    \begin{subfigure}[b]{0.15\textwidth}
        \centering
        \includegraphics[width=\textwidth]{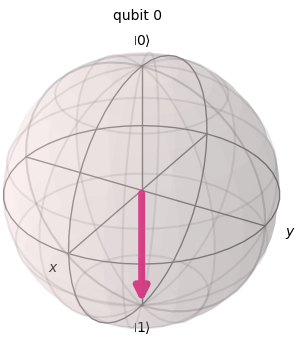}
        \caption{$d_3$ in $P_{I}$}
        \label{fig:d1_I}
    \end{subfigure}
     \hfill
    \begin{subfigure}[b]{0.30\textwidth}
        \centering
        \includegraphics[width=\textwidth]{images/z00.png}
        \caption{$z_1, z_2$ in $P_{A1}$}
        \label{fig:z00_pa1}
    \end{subfigure}
    \vspace{-5 pt}
    \caption{The flow of data and ancillary $q$-tokens in SISO quantum buffer example.}
    \label{fig:siso_bs}
\end{figure}

\vspace{-5pt}\noindent
We implemented the SISO quantum buffer that was described in Figure~\ref{fig:siso} with three data $q$-tokens ($n = 3$) $d_1,d_2,d_3$ in $P_I$ and two ancillary $q$-tokens ($m = 2$) $z_1,z_2,$ in $P_A$ where $a_1= |10\rangle ,  a_2 = |1\rangle, a_3 = |1\rangle$ and $z_1 = z_2 = |0\rangle$ as shown in Figure~\ref{fig:siso_bs}(a) and~\ref{fig:siso_bs}(b). Initially, $T_1$ is enabled and when its fires, it forwards the data $q$-token $d_1$ to $P_O$ and ancillary $q$-token $z_1$ to $P_{A1}$ as shown in Figure~\ref{fig:siso_bs}(c) and Figure~\ref{fig:siso_bs}(d). Again, $T_1$ is enabled and when its fires, it forwards the data $q$-token $d_2$ to $P_O$ and ancillary $q$-token $z_2$ to $P_{A1}$ as shown in Figure~\ref{fig:siso_bs}(g) and~\ref{fig:siso_bs}(i). The data $q$-token $d_3$ will remain in $P_I$ as there is no ancillary $q$-token left in $P_A$ to fire $T_1$ as shown in Figure~\ref{fig:siso_bs}(h).  We note that to process two data $q$-tokens, two transitions are fired.

\vspace{-10pt}
\subsection*{B2. The SIMO QPN Simulations}

\begin{figure}[!ht]
    \centering
    \begin{subfigure}[b]{0.52\textwidth}
        \centering \includegraphics[width=\textwidth]{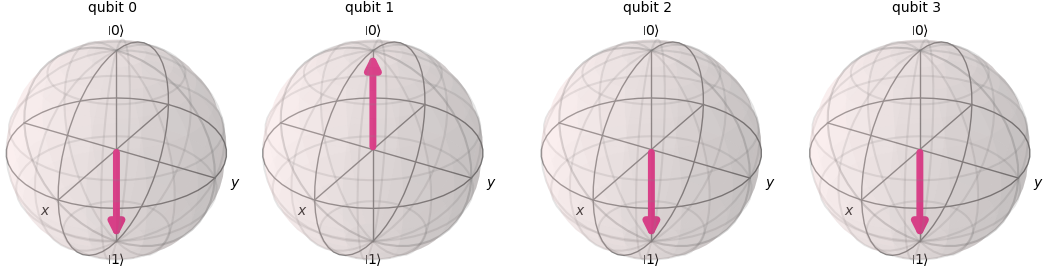}
        \caption{$d_1, d_2, d_3, d_4$ in $P_I$}
        \label{fig:simo_d1011}
    \end{subfigure}
    \hfill
    \begin{subfigure}[b]{0.39\textwidth}
        \centering \includegraphics[width=\textwidth]{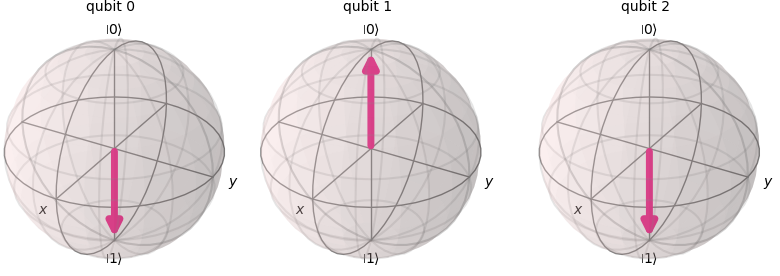}
        \caption{$z_1, z_2, z_3$ in $P_A$}
        \label{fig:simo_z101}
    \end{subfigure}
    \hfill
    \begin{subfigure}[b]{0.26\textwidth}
        \centering
        \includegraphics[width=\textwidth]{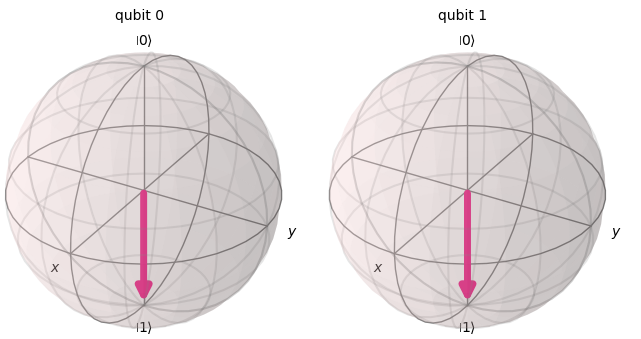}
        \caption{$d_1, d_3$ in $P_{O2}$}
        \label{fig:simo_d11}
    \end{subfigure}
    \hfill
    \begin{subfigure}[b]{0.13\textwidth}
        \centering
        \includegraphics[width=\textwidth]{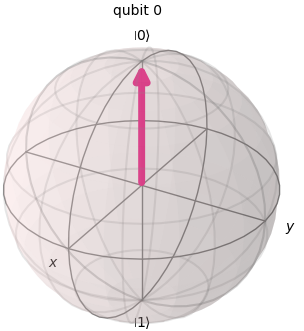}
        \caption{$d_2$ in $P_{O1}$}
        \label{fig:simo_d2}
    \end{subfigure}
    \hfill
    \begin{subfigure}[b]{0.13\textwidth}
        \centering
        \includegraphics[width=\textwidth]{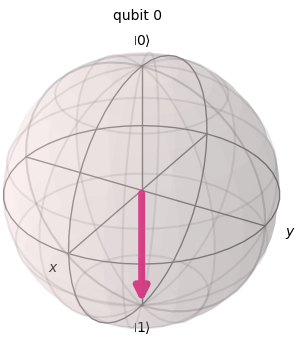}
        \caption{$d_4$ in $P_I$}
        \label{fig:simo_d4}
    \end{subfigure}
    \hfill
    \begin{subfigure}[b]{0.39\textwidth}
        \centering
        \includegraphics[width=\textwidth]{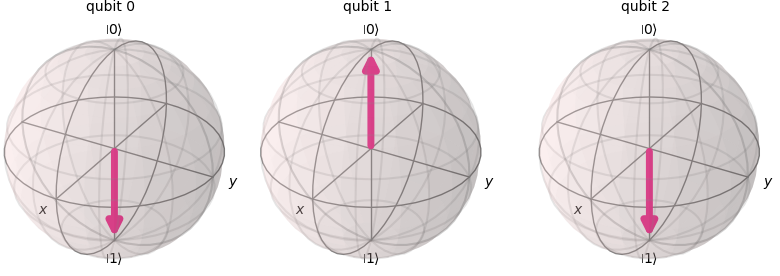}
        \caption{$z_1, z_2, z_3$ in $P_{A1}$}
        \label{fig:simo_z101b}
    \end{subfigure}
    \vspace{-5 pt}
    
    \caption{The flow of data and ancillary $q$-tokens in SIMO quantum buffer example.}
    \label{fig:simo_bs}
\end{figure}

\vspace{-5pt}\noindent
We implemented the SIMO quantum buffer that was described in Figure~\ref{fig:simo} with two output places $P_{O1}, P_{O2}$, two transitions $T_1,T_2$, four data $q$-tokens ($n = 4$) $d_1,d_2,d_3,d_4$ in $P_I$ and three ancillary $q$-tokens ($m = 3$) $z_1,z_2,z_3$ in $P_A$ where $d_1=d_3=d_4= |1\rangle,  d_2=|0\rangle$ and $z_1=z_3=|1\rangle, z_2=|0\rangle$ as shown in Figure~\ref{fig:simo_bs}(a) and~\ref{fig:simo_bs}(b). Initially, the first ancillary $q$-token $z_1=|1\rangle$, which enables the transition $T_2$ and when its fires, it forwards the data and ancillary $q$-token $d_1$ and $z_1$ to $P_{O2}$ and $P_{A1}$. The next ancillary $q$-token $z_2=|0\rangle$, which enables the transition $T_1$ and when its fires, it forwards the data and ancillary $q$-token $d_2$ and $z_2$ to $P_{O1}$ and $P_{A1}$. Lastly,  the ancillary $q$-token $z_1=|1\rangle$, which enables the transition $T_2$ and when its fires, it forwards the data and ancillary $q$-token $d_3$ and $z_3$ to $P_{O2}$ and $P_{A1}$ as shown in Figure~\ref{fig:simo_bs}(c) and~\ref{fig:simo_bs}(f), whereas data $q$-token $d_4$ will remain in $P_I$ as there is no ancillary $q$-token left in $P_A$ as shown in Figure~\ref{fig:simo_bs}(e).

\vspace{15 pt}

\subsection*{B3. The MISO QPN Simulations}
\begin{figure}[!ht]
    \centering
    \begin{subfigure}[b]{0.26\textwidth}
        \centering
        \includegraphics[width=\textwidth]{images/d10.png}
        \caption{$d_1, d_2$ in $P_{I1}$}
        \label{fig:mi1_d10}
    \end{subfigure}
    \hfill
    \begin{subfigure}[b]{0.13\textwidth}
        \centering
        \includegraphics[width=\textwidth]{images/d1.png}
        \caption{$d_3$ in $P_{I2}$}
        \label{fig:mi2_d1}
    \end{subfigure}
    \hfill
    \begin{subfigure}[b]{0.26\textwidth}
        \centering
        \includegraphics[width=\textwidth]{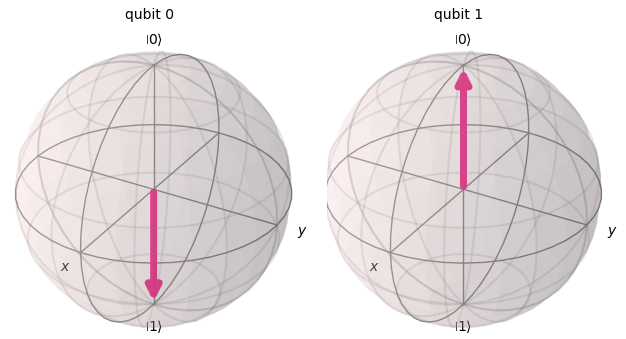}
        \caption{$z_1, z_2$ in $P_{A}$}
        \label{fig:ma_z10}
    \end{subfigure}

    \vspace{5pt}
    \begin{subfigure}[b]{0.13\textwidth}
        \centering
        \includegraphics[width=\textwidth]{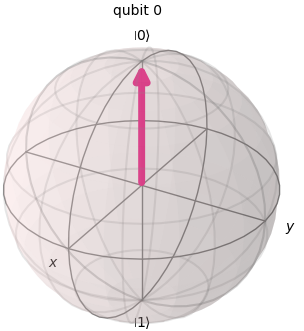}
        \caption{$d_2$ in $P_{I1}$}
        \label{fig:mi2_d0}
    \end{subfigure}
    \hfill
    \begin{subfigure}[b]{0.26\textwidth}
        \centering
        \includegraphics[width=\textwidth]{images/d11.png}
        \caption{$d_3, d_1$ in $P_O$}
        \label{fig:mo_d11}
    \end{subfigure}
    \hfill
    \begin{subfigure}[b]{0.26\textwidth}
        \centering
        \includegraphics[width=\textwidth]{images/z10.png}
        \caption{$z_1, z_2$ in $P_{A1}$}
        \label{fig:ma1_z10}
    \end{subfigure}
    
    \caption{A partial flow of data and ancillary $q$-tokens in a MISO quantum buffer example.}
    \label{fig:miso_bs}
\end{figure}

We implemented the MISO quantum buffer that was described in Figure~\ref{fig:miso} with two input places $P_{I1}, P_{I2}$, two transitions $T_1,T_2$, three data $q$-tokens ($n = 3$) $d_1,d_2$ in $P_{I1}$ ,$d_3$ in $P_{I2}$ and two ancillary $q$-tokens ($m = 3$) $z_1,z_2,z_3$ in $P_A$ where $d_1=d_3= |1\rangle,  d_2=|0\rangle$ and $z_1=|1\rangle, z_2=|0\rangle $ as shown in Figure~\ref{fig:miso_bs}(a), ~\ref{fig:miso_bs}(b) and~\ref{fig:miso_bs}(c). Initially, the first ancillary $q$-token $z_1=|1\rangle$, which enables the transition $T_2$ and when its fires, it forwards the data and ancillary $q$-token $d_3$ and $z_1$ to $P_{O}$ and $P_{A1}$. The next ancillary $q$-token $z_2=|0\rangle$, which enables the transition $T_1$ and when its fires, it forwards the data and ancillary $q$-token $d_1$ and $z_2$ to $P_{O}$ and $P_{A1}$  as shown in Figure~\ref{fig:miso_bs}(e) and~\ref{fig:simo_bs}(f), whereas data $q$-token $d_2$ will remain in $P_{I1}$ as shown in Figure~\ref{fig:miso}(d).

\vspace{15 pt}

\subsection*{B4. The MIMO QPN Simulations}

\begin{figure}[!ht]
    \centering
    \begin{subfigure}[b]{0.26\textwidth}
        \centering
        \includegraphics[width=\textwidth]{images/d10.png}
        \caption{$d_1, d_2$ in $P_{I1}$}
        \label{fig:m_d11}
    \end{subfigure}
    \hfill
    \begin{subfigure}[b]{0.26\textwidth}
        \centering
        \includegraphics[width=\textwidth]{images/d11.png}
        \caption{$d_3, d_4$ in $P_{I2}$}
        \label{fig:m_d1}
    \end{subfigure}
    \hfill
    \begin{subfigure}[b]{0.26\textwidth}
        \centering
        \includegraphics[width=\textwidth]{images/z10.png}
        \caption{$w_1, w_2$ in $P_{A}$}
        \label{fig:m_z000_0}
    \end{subfigure}
     \vspace{5pt}
    \begin{subfigure}[b]{0.26\textwidth}
        \centering
        \includegraphics[width=\textwidth]{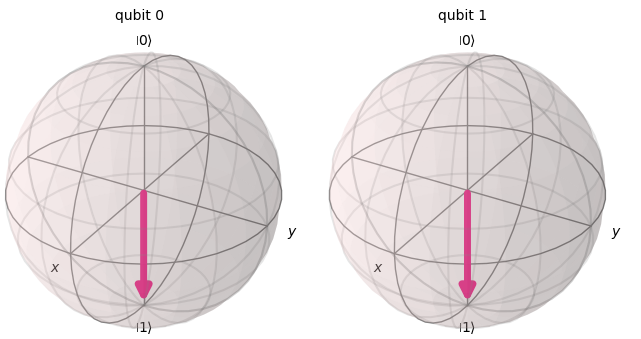}
        \caption{$z_1, z_2$ in $P_{A1}$}
        \label{fig:m_z000_1}
    \end{subfigure}
    \hfill
    \begin{subfigure}[b]{0.52\textwidth}
        \centering
        \includegraphics[width=\textwidth]{images/d1011.png}
        \caption{$w1, w2, d_3, d_1$ in $P_{DA}$}
        \label{fig:p_d125_0}
    \end{subfigure}
     \hfill
    \begin{subfigure}[b]{0.13\textwidth}
        \centering
        \includegraphics[width=\textwidth]{images/d0.png}
        \caption{$d_2$ in $P_{I1}$}
        \label{fig:p_d125_1}
    \end{subfigure}
   
    \vspace{5pt}
     \begin{subfigure}[b]{0.13\textwidth}
        \centering
        \includegraphics[width=\textwidth]{images/d1.png}
        \caption{$d_4$ in $P_{I2}$}
        \label{fig:p_d125_2}
    \end{subfigure}
    \hfill
    \begin{subfigure}[b]{0.26\textwidth}
        \centering
        \includegraphics[width=\textwidth]{images/d11.png}
        \caption{$d_3, d_1$ in $P_{O2}$}
        \label{fig:m_zA1000}
    \end{subfigure}
    \hfill
    \begin{subfigure}[b]{0.52\textwidth}
        \centering
        \includegraphics[width=\textwidth]{images/d1011.png}
        \caption{$w_1, w_2, z_1, z_2$ in $P_{A2}$}
        \label{fig:p_d125_4}
    \end{subfigure}
    
    \caption{A partial flow of data and ancillary $q$-tokens in a MIMO quantum buffer example.}
    \label{fig:mimo_bs}
\end{figure}

\vspace{-5pt}
We implemented the MIMO quantum buffer that was described in Figure~\ref{fig:mimo} with two input places $P_{I1}, P_{I2}$, four transitions $T_1,T_2,T_3,T_4$, three data $q$-tokens ($n = 4$) $d_1,d_2$ in $P_{I1}$ ,$d_3,d_4$ in $P_{I2}$ and four ancillary $q$-tokens ($m = 2$) $w_1,w_2,z_1,z_2$ where $d_1=d_3=d_4 |1\rangle,  d_2=|0\rangle$ and $w_1=z_1=z_2=|1\rangle, w_2=|0\rangle$ as shown in Figure~\ref{fig:mimo_bs}(a), ~\ref{fig:mimo_bs}(b), ~\ref{fig:mimo_bs}(c)  and~\ref{fig:mimo_bs}(d). Initially, the first ancillary $q$-token $w_1=|1\rangle$, which enables the transition $T_2$ and when its fires, it forwards the data and ancillary $q$-token $d_3$ and $w_1$ to $P_{DA}$, whereas the second ancillary $q$-token $w_2=|0\rangle$, which enables the transition $T_1$ and when its fires, it forwards the data and ancillary $q$-token $d_1$ and $w_2$ to $P_{DA}$ as shown in Figure~\ref{fig:mimo_bs}(e) . The ancillary $q$-tokens in $P_{A1}$ $z_1=z_2=|1\rangle$, which enables the transition $T_4$ and when its fires, it forwards the data and ancillary $q$-token $d_3,d_1$ and $w_1,w_2, z_1, z_2$ to $P_{O}$ and $P_{A2}$  as shown in Figure~\ref{fig:mimo_bs}(h) and~\ref{fig:mimo_bs}(i), whereas data $q$-token $d_2$ and $d_4$  will remain in $P_{I1}$ and $P_{I2}$ as shown in Figure~\ref{fig:mimo_bs}(f), ~\ref{fig:mimo_bs}(g) .

\vspace{15 pt}

\vspace{-35pt}

\subsection*{B5.  Priority QPN}

\begin{figure}[!ht]
    \centering
    \begin{subfigure}[b]{0.12\textwidth}
        \centering \includegraphics[width=\textwidth]{images/z0.png}
        \caption{$d_1$ in $P_{I1}$}
        \label{fig:p_d1}
    \end{subfigure}
    \hfill
    \begin{subfigure}[b]{0.24\textwidth}
        \centering \includegraphics[width=\textwidth]{images/z00.png}
        \caption{$w_1,w_2$ in $P_A$}
        \label{fig:p_z1}
    \end{subfigure}
    \hfill
    \begin{subfigure}[b]{0.24\textwidth}
        \centering
        \includegraphics[width=\textwidth]{images/d11.png}
        \caption{$d_2, d_3$ in $P_{I2}$}
        \label{fig:p_d23}
    \end{subfigure}
    \hfill
    \begin{subfigure}[b]{0.24\textwidth}
        \centering
        \includegraphics[width=\textwidth]{images/z00.png}
        \caption{$z_1,z_2$ in $P_{A1}$}
        \label{fig:p_z23}
    \end{subfigure}
    \hfill
    \begin{subfigure}[b]{0.36\textwidth}
        \centering
        \includegraphics[width=\textwidth]{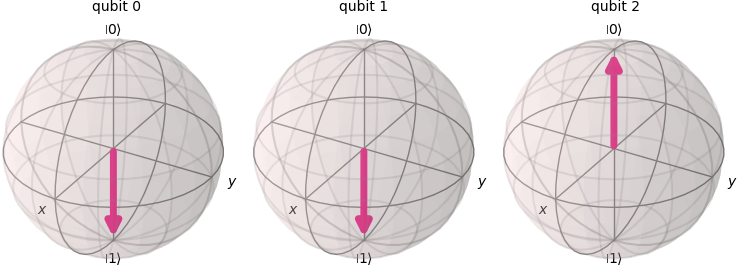}        \caption{$d_2,d_3,d_1$ in $P_O$}
        \label{fig:p_d123}
    \end{subfigure}
    \hfill
    \begin{subfigure}[b]{0.12\textwidth}
        \centering
        \includegraphics[width=\textwidth]{images/z0.png}
        \caption{$w_2$ in $P_{A}$}
        \label{fig:p_z123a}
    \end{subfigure}
    \hfill
    \begin{subfigure}[b]{0.36\textwidth}
        \centering
        \includegraphics[width=\textwidth]{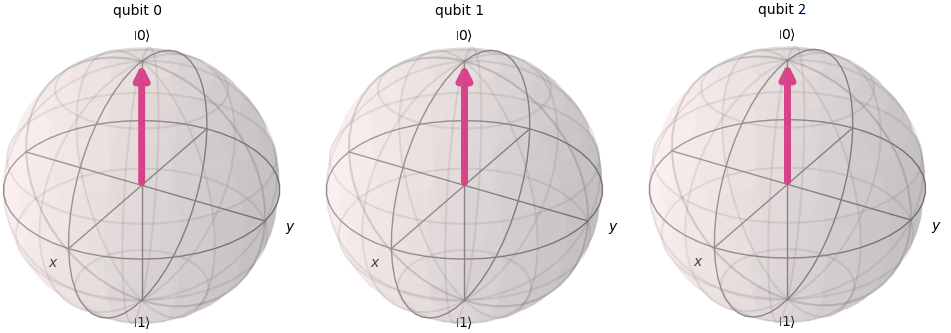}
        \caption{$w_1, z_1, z_2$ in $P_{A2}$}
        \label{fig:p_z123}
    \end{subfigure}
    \vspace{-5 pt}
    
    \caption{A partial flow of data and ancillary $q$-tokens in priority quantum buffer example.}
    \label{fig:priority_bs}
\end{figure}

\vspace{-5pt}
We implement the priority quantum buffer that was described in Figure~\ref{fig:priority}  with three data $q$-tokens ($n = 3$) $d_1,d_2,d_3$ and four ancillary $q$-tokens ($m = 2$) $w_1,w_2,z_1,z_2$ where $d_1= |0\rangle, d_2=d_3=|1\rangle$ and $w_1=w_2=z_1=z_2=|0\rangle$. The data and ancillary $q$-tokens $d_1$ and $w_1,w2$ are placed in low priority group $P_{I1}$ and $P_{A}$ as shown in Figure~\ref{fig:priority_bs} (a) and~\ref{fig:priority_bs} (b), whereas data and ancillary $q$-tokens $d_2,d_3$ and $z_1,z_2$ are placed in high priority group $P_{I2}$ and $P_{A1}$ as shown in Figure~\ref{fig:priority_bs}(c) and~\ref{fig:priority_bs}(d). Initially, the enabled transitions are $T_1$ and $T_2$, where $T_1$ will fire once and $T_2$ will fire twice, when $T_1$ fires, it forwards the data and ancillary $q$-token $d_1$ and $w_1$ to $P_{DA1}$, when $T_2$ fires, it forwards the data and ancillary $q$-token $d_2,d_3$ and $z_1,z_2$ to $P_{DA2}$. If $P_{DA2}$ is empty, then $T_3$ will fire, otherwise $T_4$ fires. When $T_3$ fires, it forwards the data $q$-token $d_1$ to $P_O$ and ancillary $q$-token $w_1$ to $P_{A2}$, and when $T_4$ fires, it forwards the data $q$-token $d_2,d_3$ to $P_O$ and ancillary $q$-token $z_1,z_2$ to $P_{A2}$ as shown in Figure~\ref{fig:priority_bs}(e) and~\ref{fig:priority_bs} (g), whereas ancillary $q$-token $w_2$ will remain in $P_A$ as there is no data $q$-token left in $P_{I1}$ as shown in Figure~\ref{fig:priority_bs}(f). 

\vspace{15 pt}

\vspace{10pt}
\section{Concluding Remarks}
\label{conclusion}

\vspace{-8pt}
This paper simplified the quantum Petri net (QPN) model  given in~\cite{Papavarnavas-2021} and used it to introduce quantum buffer designs to route quantum tokens between a set of input places and a set of output places. It presented a quantum S-R flip-flop to design quantum registers to hold quantum tokens both at input and output places and the places within the proposed quantum buffer designs. Ancillary $q$-tokens have been used in all five quantum buffer designs to fire transitions in order to route data $q$-tokens to their target (output) places. In particular, ancillary $q$-tokens were used as addresses either to select data $q$-tokens from the input places (MISO quantum buffer) or target output places to transfer $q$-tokens (SIMO quantum buffer) or both (MIMO quantum buffer). $\!\!\!\!$ Quantum buffers presented here are generalizations of classical buffers and as such they are expected to play a crucial role in quantum computers and networks much the same way classical buffers do in classical computers and networks. Thus, each type of quantum buffer plays a similar role in a quantum system as its counterpart does in the classical domain. For example, a quantum priority buffer may be used to prioritize the processing of some $q$-tokens over others because of the urgency of service they would need.  

\vspace{-9pt}
The quantum buffer designs have been validated using the IBM composer and quantum computers as well as  Qiskit  and Python to simulate the quantum buffer designs. It is noted that in the MIMO quantum buffer design, the placement of ancillary $q$-tokens into separate places forces a serialization of the transfer of data $q$-tokens from input places to output places.  This is done for the clarity of presentation of the MIMO quantum buffer design. The serialization can easily be removed by integrating the ancillary $q$-tokens to the data $q$-tokens within the input places, in which case data $q$ tokens can be transferred from input places to output places in parallel according to any assignment between the input and output places. As for future work, it will be worthwhile to validate our quantum buffer designs on a quantum computer as our validations were carried out in Qiskit only. Another direction of investigation is to look into the use of our quantum buffer designs as a basis for constructing quantum packet switching networks~\cite{shukla,ratan}. These and other further investigations will be deferred to another place.

\section{Appendix}

Here we list the Open QASM and Qiskit code we used to test the quantum S-R-flip-flop and quantum buffer designs given in the paper.
\subsection{Q-S-R Circuit 1 OpenQasm Script}
\begin{lstlisting}[caption={Q-S-R Circuit 1 Script}, label={lst:qasm_1}]
OPENQASM 2.0;
include "qelib1.inc";
qreg q[6];
creg c[6];
// Labels q[0]=S, q[1]=R, q[3]=QP, q[4]=Q
// Labels Ancillary qubits q[2], q[5],q[6]
// Initialization
barrier q[0], q[1], q[2], q[3], q[4], q[5];
//Logic Circuit
ccx q[1], q[0], q[2];
x q[1];
x q[0];
cx q[0], q[2];
cx q[1], q[3];
x q[1];
//barrier q[0], q[1], q[2], q[3], q[4], q[5];
c3x q[2], q[0], q[1], q[5];
c3x q[5], q[0], q[1], q[2];
c3x q[2], q[0], q[1], q[5];
c3x q[3], q[0], q[1], q[4];
c3x q[4], q[0], q[1], q[3];
c3x q[3], q[0], q[1], q[4];
//barrier q[0], q[1], q[2], q[3], q[4], q[5];
x q[1];
x q[0];
c3x q[3], q[0], q[1], q[5];
c3x q[5], q[0], q[1], q[3];
c3x q[3], q[0], q[1], q[5];
c3x q[2], q[0], q[1], q[4];
c3x q[4], q[0], q[1], q[2];
c3x q[2], q[0], q[1], q[4];
x q[1];
barrier q[0], q[1], q[2], q[3], q[4], q[5];
measure q[0] -> c[0];
measure q[1] -> c[1];
measure q[2] -> c[2];
measure q[3] -> c[3];
measure q[4] -> c[4];
measure q[5] -> c[5];
\end{lstlisting}
\subsection{Q-S-R Circuit 2 OpenQasm Script}
\begin{lstlisting}[caption={Q-S-R Circuit 2 Script}, label={lst:qasm_2}]
OPENQASM 2.0;
include "qelib1.inc";
qreg q[6];
creg c[6];
x q[0];
x q[2];
ccx q[0], q[2], q[3];
cx q[3], q[4];
x q[0];
x q[2];
//barrier q[0], q[1], q[2], q[3], q[4], q[5];
x q[1];
x q[3];
ccx q[1], q[3], q[5];
x q[1];
x q[3];
//barrier q[0], q[1], q[2], q[3], q[4], q[5];
ccx q[1], q[2], q[4];
barrier q[0], q[1], q[2], q[3], q[4], q[5];
measure q[0] -> c[0];
measure q[1] -> c[1];
measure q[2] -> c[2];
measure q[3] -> c[3];
measure q[4] -> c[4];
measure q[5] -> c[5];
\end{lstlisting}
\subsection{Q-J-K Circuit  OpenQasm Script}
\begin{lstlisting}[caption={Q-J-K Circuit Script}, label={lst:qasm_3}]
OPENQASM 2.0;
include "qelib1.inc";
qreg q[8];
creg c[8];
cx q[0], q[3];
x q[0];
cx q[0], q[2];
ccx q[2], q[1], q[0];
ccx q[1], q[2], q[3];
cx q[1], q[0];
cx q[1], q[2];
cx q[2], q[0];
cx q[2], q[3];
x q[1];
cx q[0], q[5];
x q[3];
x q[1];
cx q[0], q[1];
ccx q[1], q[4], q[0];
ccx q[1], q[4], q[5];
cx q[4], q[0];
cx q[4], q[1];
cx q[1], q[0];
x q[4];
cx q[1], q[5];
cx q[0], q[6];
cx q[0], q[7];
x q[5];
barrier q[0], q[1], q[2], q[3], q[4], q[5], q[6], q[7];
measure q[0] -> c[0];
measure q[1] -> c[1];
measure q[2] -> c[2];
measure q[3] -> c[3];
measure q[4] -> c[4];
measure q[5] -> c[5];
measure q[6] -> c[6];
measure q[7] -> c[7];
\end{listing}
\subsection{Imports}
\begin{lstlisting}[caption={Imports}, label={lst:imports}]
import random
import string
from qiskit import *
from qiskit.quantum_info import Statevector
from qiskit.visualization import plot_bloch_multivector
import numpy as np
\end{lstlisting}
\subsection{Token Class}
\begin{lstlisting}[caption={Token Class}, label={lst:token}]
class Token:
    def __init__(self, name, quantum_state="0"):
        self.name = name
    def __str__(self):
        return f"{self.name}"
\end{lstlisting}

\subsection{Place Class}
\begin{lstlisting}[caption={Place Class}, label={lst:place}]
class Place:
    def __init__(self, name):
        self.name = name
        self.tokens = []

    def add_token(self, token):
        self.tokens.append(token)

    def remove_token(self):
        if self.tokens:
            return self.tokens.pop(0)
        return None

    def token_count(self):
        return len(self.tokens)
    
    def __str__(self):
        return f"Place [name={self.name}, tokens={', '.join([str(t) for t in self.tokens])}]"
\end{lstlisting}

\subsection{Transition Class}
\begin{lstlisting}[caption={Transition Class}, label={lst:transition}]
class Transition:
    def __init__(self, name):
        self.name = name
        self.input = {}
        self.output = {}

    def can_fire(self, marking):
        for place, count in self.input.items():
            if marking.get_token_count(place) < count:
                return False
        return True

    def fire(self, marking):
        tok1 = ""
    
        if not self.can_fire(marking):
            print(f"cannot fire {self.name}")
            return

        print(f"fire {self.name}")
        tok2 = ""

        for place in self.input.keys():
            tokens = marking.get_tokens(place)
            if tokens:
                if place.name[1] == 'D':
                    name = tokens.pop(0).name
                    
                    if len(name) == 4:
                        tok2 = name[2:4]
                        tok1 = name[0:2] + tok1
                else:
                    tok1 += tokens.pop(0).name

                marking.set_tokens(place, tokens)

        tok3 = tok1
        for place in self.output.keys():
            if place.name[1] == 'D':
                tokens = marking.get_tokens(place)
                tokens.append(Token(tok1, "0"))
                marking.set_tokens(place, tokens)              
            else:
                tokens = marking.get_tokens(place)
                if not tok2:
                    tok2 = tok1[2:]
                    tok3 = tok1[0:2]
                else :
                    tok2 = tok2+tok1[2:]
                    tok3 = tok1[0:2]
                if place.name[1] == 'O':               
                    tokens.append(Token(tok3, "0"))
                if place.name[1] == 'A':               
                    tokens.append(Token(tok2, "0"))
                if place.name[1] == 'D':
                    tokens.append(Token(tok1, "0"))
                marking.set_tokens(place, tokens)
        return tok2,tok1
\end{lstlisting}

\subsection{Marking Class}
\begin{lstlisting}[caption={Marking Class}, label={lst:marking}]
class Marking:
    def __init__(self):
        self.place_tokens = {}
        self.count_a = 1
        self.count_z = 1

    def initialize_tokens(self, place: Place, count: int) -> None:
        if place.name[1] == 'A':
            start_char = 'z'
            current_count = self.count_z
        else:
            start_char = 'd'
            current_count = self.count_a

        tokens = [Token(f"{start_char}{i + current_count}") for i in range(count)]
        if start_char == 'z':
            self.count_z += count
        else:
            self.count_a += count

        self.place_tokens[place] = tokens    
    
    def get_token_count(self, place):
        return len(self.place_tokens.get(place, []))

    def add_token(self, place, token):
        self.place_tokens[place] = self.place_tokens.get(place, []) + [token]
    
    def get_tokens(self, place):
        return self.place_tokens[place]

    def set_tokens(self, place, count):
        self.place_tokens[place] = count

    def remove_token(self, place):
        if self.place_tokens.get(place, []):
            return self.place_tokens[place].pop(0)
        return None
   
    def print(self):
        for place, tokens in self.place_tokens.items():
            print(f"{place.name}: {[str(token) for token in tokens]}")

\end{lstlisting}

\subsection{PetriNet Class}
\begin{lstlisting}[caption={PetriNet Class}, label={lst:petrinet}]
class PetriNet:
    def __init__(self):
        self.places = []
        self.transitions = []

    def initialize_places(self, names):
        for name in names:
            place = Place(name)
            self.places.append(place)

    def initialize_transitions(self, input_dict, output_dict, names):
        for name in names:
            transition = Transition(name)
            for place in self.places:
                if place.name in input_dict.get(name, []):
                    transition.input[place] = 1
                if place.name in output_dict.get(name, []):
                    transition.output[place] = 1
            self.transitions.append(transition)

    def initialize_marking(self, place_names, token_counts):
        marking = Marking()
        for name, count in zip(place_names, token_counts):
            for place in self.places:
                if place.name == name:
                    marking.initialize_tokens(place, count)
        return marking

    def print_enabled_transitions(self, marking):
        enabled = [t for t in self.transitions if t.can_fire(marking)]
        print("Enabled Transitions:", [t.name for t in enabled])
        return enabled
\end{lstlisting}

\subsection{SISO Script}
\begin{lstlisting}[caption={SISO Script}, label={lst:siso-script}]
if __name__ == "__main__":
        net = PetriNet()
        places = ["PI", "PA", "PA1", "PO"]
        transitions = ["T1"]
        input_dict = {
            "T1": ["PI", "PA"]
        }
        output_dict = {
            "T1": ["PA1", "PO"]
        }
        initial_markings = [3, 2, 0, 0]
        
        net.initialize_places(places)
        net.initialize_transitions(input_dict, output_dict, transitions)
        current_marking = net.initialize_marking(places, initial_markings)
variables = {
        'd1': [0,0,1,0], #10 state
        'd2': [0,1], #0 state
        'd3': [0,1], #1 state
        'z1': [1,0], #0 state
        'z2': [1,0], #0 state
    } 

def match_variable(name, variables):
    for var_name, var_value in variables.items():
        if name == var_name:
            return var_value
    return None

def compose_tensor_product(state1, state2):
    state1 = np.array(state1, dtype=complex)
    state2 = np.array(state2, dtype=complex)
    combined_state = np.kron(state1, state2)
    return combined_state

num_iterations =3  
current_marking.print()
for _ in range(num_iterations):
    enabled_transitions = net.print_enabled_transitions(current_marking)
    
    if enabled_transitions:
        tok1,tok2=enabled_transitions[0].fire(current_marking)
        if len(tok2) == 4:
            data = tok2[:2]
            ancil = tok2[2:]       
            d = match_variable(data, variables)
            a = match_variable(ancil, variables)
            tp=compose_tensor_product(d,a)
            plot_bloch_multivector(tp)
        
        current_marking.print()
            
    else:
        print("No enabled transitions available.")
        break  
\end{lstlisting}

\subsection{SIMO Script}
\begin{lstlisting}[caption={SIMO Script}, label={lst:simo-script}]
if __name__ == "__main__":
        net = PetriNet()

        places = ["Pin", "PA", "PA1", "PO1", "PO2"]
        transitions = ["T1", "T2"]
        input_dict = {
            "T1": ["Pin", "PA"],
            "T2": ["Pin", "PA"]
        }
        output_dict = {
            "T1": ["PA1", "PO1"],
            "T2": ["PA1", "PO2"]
        }
        initial_markings = [4, 3, 0, 0, 0]
        
        net.initialize_places(places)
        net.initialize_transitions(input_dict, output_dict, transitions)
        current_marking = net.initialize_marking(places, initial_markings)
variables = {
        'd1': [0,1], #1 state
        'd2': [1,0], #0 state
        'd3': [0,1], #1 state
        'd4': [0,1], #1 state
        'z1': [0,1], #1 state select T2
        'z2': [1,0], #0 state select T1
        'z3': [0,1]  #1 state select T2
    } 

def match_variable(name, variables):
    for var_name, var_value in variables.items():
        if name == var_name:
            return var_value
    return None

def compose_tensor_product(state1, state2):
    state1 = np.array(state1, dtype=complex)
    state2 = np.array(state2, dtype=complex)
    combined_state = np.kron(state1, state2)  
    return combined_state
num_iterations = 6  
current_marking.print()
for i in range(num_iterations):
    enabled_transitions = net.print_enabled_transitions(current_marking)
    
    if enabled_transitions:
        
        key = 'z' + str(i+1)
        if variables[key] == [1, 0]:
            index = next((i for i, transition in enumerate(enabled_transitions) if transition.name == 'T1'), None)
            print("T1 in enabled transition",index)
        if variables[key] == [0, 1]:
            index = next((i for i, transition in enumerate(enabled_transitions) if transition.name == 'T2'), None)
            print("T2 in enabled transition",index)
        
        
        tok1,tok2=enabled_transitions[index].fire(current_marking)
        if len(tok2) == 4:
            data = tok2[:2]
            ancil = tok2[2:]       
            d = match_variable(data, variables)
            a = match_variable(ancil, variables)
            tp=compose_tensor_product(d,a)
            plot_bloch_multivector(tp)
        
        current_marking.print()
            
    else:
        print("No enabled transitions available.")
        break  

\end{lstlisting}

\subsection{MISO Script}
\begin{lstlisting}[caption={MISO Script}, label={lst:miso-script}]
if __name__ == "__main__":
        net = PetriNet()

        places = ["Pin1","Pin2", "PA","PA1", "PO"]
        transitions = ["T1", "T2"]
        input_dict = {
            "T1": ["Pin1", "PA"],
            "T2": ["Pin2", "PA"]
        }
        output_dict = {
            "T1": ["PA1", "PO"],
            "T2": ["PA1", "PO"]
        }
        initial_markings = [3, 2, 3, 0, 0]
        
        net.initialize_places(places)
        net.initialize_transitions(input_dict, output_dict, transitions)
        current_marking = net.initialize_marking(places, initial_markings)
variables = {
        'd1': [0,1], #1 state
        'd2': [1,0], #0 state
        'd3': [0,1], #1 state
        'd4': [0,1], #1 state
        'd5': [0,1], #1 state
        'z1': [1,0], #0 state
        'z2': [0,1], #1 state
        'z3': [0,1]  #1 state
    } 

def match_variable(name, variables):
    for var_name, var_value in variables.items():
        if name == var_name:
            return var_value
    return None

def compose_tensor_product(state1, state2):
    state1 = np.array(state1, dtype=complex)
    state2 = np.array(state2, dtype=complex)
    
    combined_state = np.kron(state1, state2)
    
   
    return combined_state
num_iterations = 6  
current_marking.print()
for i in range(num_iterations):
    enabled_transitions = net.print_enabled_transitions(current_marking)
    
    if enabled_transitions:
        key = 'z' + str(i+1)
        if variables.get(key) == [1, 0]:
            index = next((i for i, transition in enumerate(enabled_transitions) if transition.name == 'T1'), None)
        elif variables.get(key) == [0, 1]:
            index = next((i for i, transition in enumerate(enabled_transitions) if transition.name == 'T2'), None)
        else:
            index = random.randint(0, len(enabled_transitions) - 1) 
            
        tok1,tok2=enabled_transitions[index].fire(current_marking)
        if len(tok2) == 4:
            data = tok2[:2]
            ancil = tok2[2:]       
            d = match_variable(data, variables)
            a = match_variable(ancil, variables)
            tp=compose_tensor_product(d,a)
            plot_bloch_multivector(tp)
        
        current_marking.print()
            
    else:
        print("No enabled transitions available.")
        break   
      
\end{lstlisting}

\subsection{MIMO Script}
\begin{lstlisting}[caption={MIMO Script}, label={lst:mimo-script}]
if __name__ == "__main__":
        net = PetriNet()

        places = ["Pin1","Pin2", "PDA", "PA1","PA2","PA3", "PO1","PO2"]
        transitions = ["T1", "T2","T3","T4"]
        input_dict = {
            "T1": ["Pin1", "PA1"],
            "T2": ["Pin2", "PA1"],
            "T3": ["PDA", "PA2"],
            "T4": ["PDA", "PA2"]
        }
        output_dict = {
            "T1": ["PDA"],
            "T2": ["PDA"],
            "T3": ["PA3", "PO1"],
            "T4": ["PA3", "PO2"]
        }
        initial_markings = [3, 2, 0, 3, 3, 0, 0, 0]
        
        net.initialize_places(places)
        net.initialize_transitions(input_dict, output_dict, transitions)
        current_marking = net.initialize_marking(places, initial_markings)
variables = {
        'd1': [0,1], #1 state
        'd2': [1,0], #0 state
        'd3': [0,1], #1 state
        'd4': [0,1], #1 state
        'd5': [0,1], #1 state
        'z1': [1,0], #0 state select T1
        'z2': [1,0], #0 state select T1
        'z3': [0,1], #1 state select T2
        'z4': [0,0,1,0], #10 state select T3
        'z5': [0,0,1,0], #10 state select T3
        'z6': [0,0,0,1] #11 state select T4
        
    } 

def match_variable(name, variables):
    for var_name, var_value in variables.items():
        if name == var_name:
            return var_value
    return None

def compose_tensor_product(state1, state2):
    state1 = np.array(state1, dtype=complex)
    state2 = np.array(state2, dtype=complex)
    combined_state = np.kron(state1, state2)
    
   
    return combined_state
num_iterations = 6  
current_marking.print()
for i in range(num_iterations):
    enabled_transitions = net.print_enabled_transitions(current_marking)
    
    if enabled_transitions:        
        
        key = 'z' + str(i+1)
        if variables.get(key) == [1,0]:
            index = next((i for i, transition in enumerate(enabled_transitions) if transition.name == 'T1'), None)
            print("T1 in enabled transition",index)
        elif variables.get(key) == [0,1]:
            index = next((i for i, transition in enumerate(enabled_transitions) if transition.name == 'T2'), None)
            print("T2 in enabled transition",index)
        elif variables.get(key) == [0,0,1,0]:
            index = next((i for i, transition in enumerate(enabled_transitions) if transition.name == 'T3'), None)
            print("T3 in enabled transition",index)
        elif variables.get(key) == [0,0,0,1]:
            index = next((i for i, transition in enumerate(enabled_transitions) if transition.name == 'T4'), None)
            print("T4 in enabled transition",index)
        else:
            index = random.randint(0, len(enabled_transitions) - 1) 
        
        tok1,tok2=enabled_transitions[index].fire(current_marking)
        if len(tok2) == 4:
            data = tok2[:2]
            ancil = tok2[2:]       
            d = match_variable(data, variables)
            a = match_variable(ancil, variables)
            tp=compose_tensor_product(d,a)
            plot_bloch_multivector(tp)
        
        current_marking.print()
            
    else:
        print("No enabled transitions available.")
        break  
       
      
\end{lstlisting}

\subsection{Priority Script}
\begin{lstlisting}[caption={Priority script}, label={lst:priority-script}]
if __name__ == "__main__":
        net = PetriNet()

        places = ["Pin1","Pin2", "PDA1", "PA1","PDA2", "PA2","PA3", "PO"]
        transitions = ["T1", "T2","T3","T4"]
        input_dict = {
            "T1": ["Pin1", "PA1"],
            "T2": ["Pin2", "PA2"],
            "T3": ["PDA1"],
            "T4": ["PDA2"]
        }
        output_dict = {
            "T1": ["PDA1"],
            "T2": ["PDA2"],
            "T3": ["PA3", "PO"],
            "T4": ["PA3", "PO"]
        }
        initial_markings = [1, 2, 0, 2, 0, 2, 0, 0]
        
        net.initialize_places(places)
        net.initialize_transitions(input_dict, output_dict, transitions)
        current_marking = net.initialize_marking(places, initial_markings)
variables = {
        'd1': [1,0], #0 state
        'd2': [0,1], #1 state
        'd3': [0,1], #1 state
        'z1': [1,0], #0 state
        'z2': [1,0], #0 state
        'z3': [1,0], #0 state
        'z4': [1,0]  #0 state
    } 

def match_variable(name, variables):
    for var_name, var_value in variables.items():
        if name == var_name:
            return var_value
    return None

def compose_tensor_product(state1, state2):
    state1 = np.array(state1, dtype=complex)
    state2 = np.array(state2, dtype=complex)
    combined_state = np.kron(state1, state2)
    
   
    return combined_state
num_iterations = 6  
current_marking.print() 

for i in range(num_iterations):
    enabled_transitions = net.print_enabled_transitions(current_marking)
    
    if enabled_transitions:
        t4 = next((i for i, transition in enumerate(enabled_transitions) if transition.name == 'T4'), None)    
        if t4:
            index=t4
        else:
            index = random.randint(0, len(enabled_transitions) - 1)
        
        
        tok1,tok2=enabled_transitions[index].fire(current_marking)
        
        if len(tok2) == 4:
            data = tok2[:2]
            ancil = tok2[2:]       
            d = match_variable(data, variables)
            a = match_variable(ancil, variables)
            tp=compose_tensor_product(d,a)
            plot_bloch_multivector(tp)
        
        current_marking.print() 
            
    else:
        print("No enabled transitions available.")
        break   
\end{lstlisting}


\begin{thebibliography}{2}
\bibitem{Lanzagorta2022} M. Lanzagorta and Jeffrey Uhlmann. Quantum computer science. Springer Nature, 2022.
\vspace{-20pt}
\bibitem{Djordjevic2021} B. I. Djordjevic, Quantum information processing, quantum computing, and quantum error correction: An engineering approach. Academic Press, 2021.
\vspace{-10pt}
\bibitem{Hassija2020} V. Hassija, et al. Present landscape of quantum computing. IET Quantum Communication 1, No. 2 (2020), pp. 42-48.
\vspace{-10pt}
\bibitem{Horowitz-2019} M. Horowitz, E. Grumbling, National Academies of Sciences, Engineering, and Medicine, et al. Quantum computing: progress and prospects. 2019. The National Academies Press.
\vspace{-20pt}
\bibitem{Wolfgang2019} W. Scherer Mathematics of quantum computing. Vol. 11. Springer International Publishing, 2019.
\vspace{-10pt}
\bibitem{Shor-1994} P. W. Shor. Algorithms for quantum computation: discrete logarithms and factoring. In Proceedings 35th annual symposium on foundations of computer science, pp. 124–134, 1994. IEEE.
\vspace{-10pt}
\bibitem{Grover-1996} L. K. Grover. A fast quantum mechanical algorithm for database search. In Proceedings of the twenty-eighth annual ACM symposium on Theory of computing, pp. 212-219. 1996.
\vspace{-22pt}
\bibitem{Rhonda-2023} A-Y, Rhonda, N. Chancellor, and P. Halffmann. NP-hard but no longer hard to solve? Using quantum computing to tackle optimization problems. Frontiers in Quantum Science and Technology 2 (2023).
\vspace{-10pt}
\bibitem{Preskill-2023} J. Preskill.  Quantum computing 40 years later. In Feynman Lectures on Computation, pp. 193-244. CRC Press, 2023.
\vspace{-10pt}
\bibitem{Humble-2021} T. S. Humble, A. McCaskey, D. I. Lyakh, M. Gowrishankar, A. Frisch, and T. Monz. Quantum computers for high-performance computing. IEEE Micro, 41(5), pp.15–23, 2021. 
\vspace{-20pt}
\bibitem{Bacon-2010} D. Bacon and W. Van Dam. Recent progress in quantum algorithms. Communications of the ACM, 53(2): pp. 84–93, 2010. ACM New York, NY, USA.
\vspace{-10pt}
\bibitem{Ramezani-2020} S. Ramezani, et.al. Machine learning algorithms in quantum computing: A survey. In 2020 International joint conference on neural networks (IJCNN), pp. 1-8. IEEE, 2020.
\vspace{-10pt}
\bibitem{Aumasson-2017}  J-P. Aumasson. The impact of quantum computing on cryptography. Computer Fraud \& Security 2017, No. 6 (2017), pp. 8-11.
\vspace{-10pt}
\bibitem{Danzig-1989} P. Danzig, Finite buffers for fast multicast." ACM SIGMETRICS Performance Evaluation Review 17, No. 1 (1989), pp. 108-117.
\vspace{-10pt}
\bibitem{Kimura-1996} T. Kimura. Optimal buffer design of an M/G/s queue with finite capacity. Stochastic Models 12, No. 1 (1996), pp. 165-180.
\vspace{-10pt}
\bibitem{Kougkas-2020} A. Kougkas, H. Devarajan, and X.-H. Sun. I/O Acceleration via Multi-Tiered Data Buffering and Prefetching. Journal of Computer Science and Technology, 35: pp. 92–120, 2020.
\vspace{-10pt}
\bibitem{Kim2001} H. S. Kim and B. S. Ness. Loss probability calculations and asymptotic analysis for finite buffer multiplexers. IEEE/ACM transactions on networking, No. 6 (2001)m pp. 755-768.
\vspace{-22pt}
\bibitem{liu2023quantum} C. Liu et. al. Quantum memory: A missing piece in quantum computing units. arXiv preprint arXiv:2309.14432, 2023.
\vspace{-10pt}
\bibitem{Peterson-1977} J. L. Peterson. Petri nets. ACM Computing Surveys (CSUR), 9(3):223–252, 1977. ACM New York, NY, USA.

\vspace{-10pt}
\bibitem{Zurawski-1994} R. Zurawski and M. Zhou. Petri nets and industrial applications: A tutorial. IEEE Transactions on industrial electronics, 41(6), pp. 567–583, 1994. 
\vspace{-10pt}
\bibitem{Zhou-1996} M. Zhou and F. DiCesare. Petri net modelling of buffers in automated manufacturing systems. IEEE Transactions on Systems, Man, and Cybernetics, Part B (Cybernetics), 26(1),  pp. 157–164, 1996. 
\vspace{-10pt}
\bibitem{Ye-2003} X. Ye, J. Zhou, and X. Song. On reachability graphs of Petri nets. Computers and Electrical Engineering, 29(2), pp. 263–272, 2003. Elsevier.
\vspace{-10pt}
\bibitem{Letia-2021} T. S. Letia, E. M. Durla-Pasca, and D. Al-Janabi. Quantum Petri Nets. In 2021 25th International Conference on System Theory, Control and Computing (ICSTCC), pp. 431–436, 2021. 
\vspace{-10pt}
\bibitem{Letia-2022} T. S. Letia, E. M. Durla-Pasca, D. Al-Janabi, and O. P. Cuibus. Development of Evolutionary Systems Based on Quantum Petri Nets. Mathematics, 10(23):4404, 2022. MDPI.
\vspace{-20pt}
\bibitem{Papavarnavas-2021} G. Papavarnavas. Integrating Quantum computing concepts in Petri nets. Bachelor's Thesis, University Of CYPRUS, 2021. Department OF Computer Science.
\vspace{-10pt}
\bibitem{Lee-2024} K. F. Lee and P. Kumar. Non-Markovian Dynamics in Fiber Delay-line Buffers. arXiv, 2024. DOI: 10.48550/arXiv.2402.00274.
\vspace{-10pt}
\bibitem{Schmidt-2021} H. W. Schmidt.  How to Bake Quantum into Your Pet Petri Nets and Have Your Net Theory Too. In Service-Oriented Computing: 15th Symposium and Summer School, SummerSOC 2021, Virtual Event, September 13–17, 2021, Proceedings 15, pp. 3-33. Springer International Publishing, 2021.
\vspace{-10pt}
\bibitem{Brylinski-2002} J.-L. Brylinski and R. Brylinski. Universal quantum gates. Mathematics of quantum computation, 79, 2002.
\vspace{-10pt}
\bibitem{Simon-2010} Simon, Christoph, et al. "Quantum memories: a review based on the European integrated project 'qubit applications (QAP)'." The European Physical Journal D 58 (2010), pp. 1-22.
\vspace{-10pt}
\bibitem{misra-2018} N. K. Misra, W. Subodh and S. Bibhash. "Design of conservative, reversible sequential logic for cost efficient emerging nano circuits with enhanced testability." Ain Shams Engineering Journal 9, no. 4 (2018): 2027-2037.
\vspace{-10pt}
\bibitem{misra-2024} N. K Misra,  K. B. Bandan  and R. K. Sankit.  Utilizing a Novel Universal Quantum Gate in the Design of Fault-Tolerant Architecture. Nano Communication Networks 39 (2024): 100482.

\vspace{-10pt}
\bibitem{IBM-Qiskit} IBM Quantum. Qiskit: An Open-source Quantum Computing Framework. Retrieved from https://www.ibm.com/quantum/qiskit.
\vspace{-10pt}
\bibitem{IBM-Composer} IBM Quantum. IBM Quantum Composer. https://quantum.ibm.com/composer.
\vspace{-10pt}
\bibitem{Jupyter} Jupyter, Project Jupyter. https://jupyter.org/. 
\vspace{-10pt}
\bibitem{shukla} M. K. Shukla and A. Y. Oru\c{c}. Multicasting in quantum switching networks. IEEE
Transactions on Computers 59, no. 6, pp. 735-747, 2010.
\vspace{-10pt}
\bibitem{ratan} R. Ratan and A. Y. Oru\c{c} Self-routing quantum sparse crossbar packet concentrators.
IEEE Transactions on Computers 60, no. 10, pp. 1390-1405, 2010.
\end{thebibliography}
\end{document}